\documentclass[lettersize,journal]{IEEEtran}
\usepackage{amsmath,amsfonts}
\usepackage{algorithmic}
\usepackage{array}
\usepackage[caption=false,font=normalsize,labelfont=sf,textfont=sf]{subfig}
\usepackage{textcomp}
\usepackage{stfloats}
\usepackage{url}
\usepackage{verbatim}
\usepackage{graphicx}
\usepackage{amsthm}
\usepackage{float}
\theoremstyle{definition}
\usepackage{stfloats}  
\usepackage{afterpage}  % 提供 \afterpage
\usepackage{placeins}
\usepackage{amsthm}
\usepackage{booktabs}
\usepackage[ruled,vlined]{algorithm2e}
\usepackage{graphicx}
\usepackage{subfig}
\usepackage{multirow}
\usepackage{makecell}
\usepackage{cite}
\usepackage{siunitx}   % 可选：数值对齐
\theoremstyle{definition}
\usepackage{cleveref}
\def\BibTeX{{\rm B\kern-.05em{\sc i\kern-.025em b}\kern-.08em
    T\kern-.1667em\lower.7ex\hbox{E}\kern-.125emX}}
\usepackage{balance}
\begin{document}
\title{Graph Signal Wiener Filtering in the Linear Canonical Domain: Theory and Method Design}
\author{Xiaopeng~Cheng and Zhichao~Zhang,~\IEEEmembership{Member,~IEEE}
\thanks{This work was supported in part by the Open Foundation of Hubei Key Laboratory of Applied Mathematics (Hubei University) under Grant HBAM202404; and in part by the Foundation of Key Laboratory of System Control and Information Processing, Ministry of Education under Grant Scip20240121. \emph{(Corresponding author: Zhichao~Zhang.)}

Xiaopeng~Cheng is with the School of Mathematics and Statistics, Nanjing University of Information Science and Technology, Nanjing 210044, China (e-mail: 18355938105@163.com).

Zhichao~Zhang is with the School of Mathematics and Statistics, Nanjing University of Information Science and Technology, Nanjing 210044, China, with the Hubei Key Laboratory of Applied Mathematics, Hubei University, Wuhan 430062, China, and also with the Key Laboratory of System Control and Information Processing, Ministry of Education, Shanghai Jiao Tong University, Shanghai 200240, China (e-mail: zzc910731@163.com).}}

\maketitle

\begin{abstract}
The graph linear canonical transform (GLCT)-based filtering methods often optimize transform parameters and filters separately, which results in high computational costs and limited stability. To address this issue, this paper proposes a trainable joint optimization framework that combines GLCT parameters and Wiener filtering into an end-to-end learning process, allowing for synergistic optimization between transform domain construction and filtering operations. The proposed method not only eliminates the cumbersome grid search required by traditional strategies but also significantly enhances the flexibility and training stability of the filtering system. Experimental results on real-world graph data show the proposed method outperforms existing methods in denoising tasks, featuring superior denoising performance, higher robustness and lower computational complexity.
\end{abstract}


\section{Introduction}
\IEEEPARstart{I}{n} social networks, transportation systems, biomolecular networks, and other irregular structures, data often reside on non-Euclidean grids, where classical signal processing methods become inapplicable. Thus, graph signal processing (GSP) has emerged as a solution to address these challenges. GSP models irregularly structured data as a graph, where nodes represent data entities, edges encode their relationships, and signal values are attached to the nodes\cite{1,2,3,4,5,6}. However, graph signals inevitably suffer from noise interference during the processes of acquisition, transmission, and storage. Practical applications often necessitate the extraction of key features from complex graph signals. The essence of these requirements lies in accurately separating the smooth signal from noise. Therefore, this process relies heavily on a filtering theory adapted to the structural characteristics of graphs\cite{7,8,9,10,11,12,13,14,15}. Classical linear filtering is bulit on the properties of Euclidean spaces. However, for graph structures, which represent a type of non-Euclidean space, the classical definition methods are difficult to transfer directly. To overcome this limitation, it is necessary to extend the concept of filtering to the graph domain and gradually establish a filtering theory framework that is adaptive to graph structures.

To adapt the filtering theory framework for graph structures, the central task is to exploit graph topology to effectively separate the smooth signal from noise. To achieve this goal, existing graph signal filtering has generally developed two complementary lines: filtering design in the transform spectral domain\cite{7,8,9,10,11} and filtering implementations in the vertex domain\cite{11,12,13,14,15}. The common foundation of both approaches is to leverage graph topology to define a ``frequency" concept that characterizes signal smoothness, thereby enabling selective enhancement or suppression of different frequency components. Transform spectral domain filter design follows a vertex-spectral-vertex pipeline: the signal is first transformed from the vertex domain to the spectral domain; a filter is then applied in the spectral domain to differentially weight spectral coefficients; the result is mapped back to the vertex domain to obtain the filtered graph signal. In contrast, vertex domain filtering operates directly in the vertex domain through localized aggregation between a node and its neighbors. For each node, a local window encompassing itself and its neighbors is defined, and an aggregation function is applied to fuse their signals, yielding the filtered output for the central node. These two families are theoretically complementary and practically distinct; the subsequent sections of this paper will primarily focus on the transform spectral domain filtering.

A cornerstone of GSP is the graph Fourier transform (GFT)\cite{3,8,16,17,18}. It extends the classical Fourier transform (FT)\cite{19,20} to graph-structured data by projecting signals onto the eigenbasis of a graph shift operator (GSO)\cite{21,22,23,24,25,26}, such as the graph Laplacian or adjacency matrix, where the corresponding eigenvalues represent graph frequency. However, the GFT only facilitates analysis in the vertex and spectral domains, and making it difficult to explore the intermediate domain between these two domains. To address this limitation, Wang et al.\cite{27} introduced the graph fractional Fourier transform (GFRFT), which incorporates a fractional order parameter to interpolate between the identity transform and the full GFT. In this framework, the GFRFT is defined by applying a fractional order to the global matrix power of the GFT matrix induced by a given GSO, thus transplanting the core idea of the classical fractional Fourier transform (FRFT)\cite{28,29,30,31,32} to the graph domain. Subsequent studies broadened the GFRFT by incorporating hyper-differential operators to generalize fractional powers and ensure differentiability with respect to the fractional parameter\cite{33}. This development is significant because it enables gradient-based optimization over the fractional order. This feature makes the transformation not only flexible but also learnable, enabling it to adapt to different learning scenarios.  

Nevertheless, the GFRFT is inherently constrained by a single fractional order parameter, offering limited degrees of freedom. It is challenging to optimize the energy aggregation of different local structures in the fractional domain, and signal parameter rotation lacks sufficient granularity for modulating spectral morphology. To overcome these limitations, this paper systematically introduces the graph linear canonical transform (GLCT)\cite{34,35,36,37}. In classical signal processing, the linear canonical transform (LCT)\cite{38,39,40,41,42} generalizes rotations to an affine transformation, thereby extending both the FT and the FRFT into a more extensible parameterized linear integral transform. Compared to FT and FRFT, the LCT incorporates three independent parameters and exhibits affine transformations that include scaling and shearing operations. Thus, it is necessary to extend the LCT to the domain of GSP. Zhang et al.\cite{34,36} introduced a definition of the GLCT based on centered discrete Hermite-Gaussian functions (CDDHFs). Their approach combines graph chirp multiplication (GCM), graph scaling transform (GST), and GFRFT. Building on this direction, Li et al.\cite{35} proposed a CM-CC-CM realization that composes GCM, graph chirp convolution (GCC), and GCM. In contrast to the above two classes of GLCT based on the eigenbasis of the adjacency matrix, Chen et al.\cite{37} started from the eigenbasis of the graph Laplacian. Furthermore, they built upon CDDHFs by defining GLCT through a combination of GCM, GST, and GFRFT. Further extending the GLCT framework, Chen et al.\cite{43} later proposed a GLCT framework based on hyper-differential operators (OGLCT). This approach involves deriving hyper-differential operators associated with the GFT matrix by solving a Sylvester equation\cite{33,44,45}. Using these operators, definitions for GFRFT, GST, and GCM are formalized, ultimately yielding a learnable GLCT framework.

Among the three transforms discussed above, the GLCT offers greater degrees of freedom for structural representation in the transform spectral domain. Centered around these transform spectral domains, the idea of graph Wiener filtering (GWF)\cite{33,46,47,48,49} is to seek a transform spectral domain under which the energy of the observed signal is concentrated as much as possible, while noise is scattered as much as possible. A diagonalized optimal linear estimator is then applied in the transform spectral domain to minimize the mean squared error (MSE). In many graph tasks, signal energy and noise are not strictly restricted to low and high frequencies; instead, they are more separable in a transform spectral domain. Therefore, Koc et al.\cite{46} extended optimal GWF to the GFRFT domain, leveraging the additional degree of freedom provided by the fractional order to achieve a lower MSE. Yet for graph signals with multi-scale structure and nonstationary perturbations, a single rotational parameter is often insufficient; some structures are separated in one fractional domain while others remain coupled. To overcome this limitation, several studies extend filtering from the GFRFT to the GLCT domain. Chen et al.\cite{37} formulated filtering in the linear canonical domain and integrated it with machine learning by proposing an SGD-based scheme that optimizes only the filter parameters, which can better accommodate complex models and datasets. A complementary line of work, based on hyper-differential operators, casts the GLCT as an exponential operator and introduces a trainable OGLCT that optimizes only the transform parameters\cite{43}. Despite the flexibility of the GLCT, optimizing only the filter or only the transform still requires grid-based parameter selection, which incurs a high computational cost and limited stability.

To this end, we propose a jointly trainable GLCT-GWF framework: under a unified MSE objective, we jointly optimize the GLCT parameters and the filter coefficients. This joint learning approach solves the tasks of “choosing an appropriate transform spectral domain” and “performing optimal linear estimation within that domain” in a single step, markedly reducing the cost of grid search and delivering more robust performance. 

The main contributions of this paper are summarized as follows:
\begin{itemize}
     \item We propose a new Laplacian eigenbasis-based CM-CC-CM-GLCT to address the gap in existing CM-CC-CM-GLCT, and organize CDDHFs-GLCT and CM-CC-CM-GLCT, each covering eigenbasis expansion for both weighted adjacency and Laplacian matrices.
    \item We prove the differentiability of the core modules of GLCT under both weighted adjacency matrices and Laplacian matrices, providing theoretical support for the end-to-end optimization of transformation parameters and filter coefficients and enabling adaptive parameter adjustment.
    \item We construct a GLCT-GWF framework for the end-to-end joint optimization of GLCT parameters and filter coefficients, and verify its effectiveness and robustness across varying noise levels and graph structures in real-world graph signal denoising tasks.
\end{itemize}

The remainder of this paper is organized as follows. Section $\mathrm{II}$ provides the necessary preliminaries. Section $\mathrm{III}$ introduces the definition and theoretical properties of the proposed CM-CC-CM-GLCT.  Section $\mathrm{IV}$ presents the GLCT-GWF framework and theoretical analysis. Section $\mathrm{V}$ provides some the experimental results, followed by
the concluding remarks in Section $\mathrm{VI}$. All the technical proofs of our theoretical results are relegated to the Appendix parts.

\section{Preliminaries}
\subsection{GFT}
Let $\mathcal{G}=\{\mathcal{N},\mathcal{E},\mathbf{A}\}$ be a graph with vertex set $\mathcal{N}$, edge set $\mathcal{E}$ and adjacency matrix $\mathbf{A}\in\mathbb{R}^{N\times N}$. The weighted adjacency matrix $\mathbf{W}$ encodes edge weights and the Laplacian matrix is defined as $\mathbf{L} = \mathbf{D}-\mathbf{W}$, where $\mathbf{D}$ is the diagonal degree matrix of the graph $\mathcal{G}$. The $i$-th diagonal element $d_i$ of $\mathbf{D}$ equals the sum of the weights of all edges connected to vertex $i$. A column of signals $\mathbf{f} = [f(0), f(1), \ldots, f(N-1)]^{\mathrm{T}} \in\mathbb{R}^{N}$ defined on the graph is a mapping from the vertex set $\mathcal{N}$ to the real number $\mathbb{R}$.

In GSP, the weighted adjacency matrix $\mathbf{W}$ and the Laplacian matrix $\mathbf{L}$ are all special cases of the GSOs, denoted as $\mathbf{X}$. Let the Jordan decomposition of the GSO be written as $\mathbf{X} = \mathbf{U_X}\mathbf{\Lambda}\mathbf{U}^{-1}_\mathbf{X}$, where $\mathbf{\Lambda}$ is the Jordan block form matrix, and $\mathbf{U_X}$ contains the generalized eigenvectors of $\mathbf{X}$ in its columns.

The GFT of a signal $\mathbf{f}$ is defined as 
\begin{equation}
\hat{\mathbf{f}} = \mathcal{F}_\mathbf{X}\mathbf{f} = \mathbf{U}^{-1}_\mathbf{X}\mathbf{f},
\end{equation}
the inverse GFT is given by 
\begin{equation}
    \mathbf{f} = \mathcal{F}^{-1}_\mathbf{X}\hat{\mathbf{f}} = \mathbf{U_X}\hat{\mathbf{f}}.
\end{equation}

\subsection{GFRFT}
The GFRFT is constructed via the Jordan decomposition of the GFT matrix. Let the GFT matrix be decomposed as
\begin{equation}
    \mathcal{F}_\mathbf{X} = \mathbf{U}^{-1}_\mathbf{X} = \mathbf{P_X}\mathbf{J_X}\mathbf{P}^{-1}_\mathbf{X}.
\end{equation}
Then, the GFRFT matrix of order $\alpha$ is defined as
\begin{equation}
    \mathcal{F}^{\alpha}_\mathbf{X} = \mathbf{P_X}\mathbf{J}^{\alpha}_\mathbf{X}\mathbf{P}^{-1}_\mathbf{X}.
\end{equation}
The definition of the $\alpha$-th order GFRFT is given by
\begin{equation}
    \hat{\mathbf{f}}_\alpha = \mathcal{F}^{\alpha}_\mathbf{X}\mathbf{f}. 
    \label{eq:gfrgt}
\end{equation}

The GFRFT preserves the additivity property, which enables the definition of the inverse transform
\begin{equation}
    \mathbf{f} = \mathcal{F}^{-\alpha}_\mathbf{X}\hat{\mathbf{f}}_\alpha.
\end{equation}
When $\alpha = 0$, the GFRFT reduces to the identity matrix, i.e., $\mathcal{F}^{0}_\mathbf{X} = \mathbf{I}_N$ and reduces to the GFT when $\alpha = 1$, i.e., $\mathcal{F}^{1}_\mathbf{X} = \mathcal{F}_\mathbf{X}$.
\subsection{GST}
The GST serves as a fundamental multi-scale analysis operation in GSP. Its primary function is to achieve compression or stretching of signal frequency components by applying scaling transformations to the spectral domain representation of the graph structure or graph signals. Depending on the graph matrix used, the graph scaling transform can be implemented in the following two main ways:
\subsubsection{GST based on the Weighted Adjacency Matrix}
Define $\mathbf{S_W} = \frac{1}{\sigma}\mathbf{W}$ as the graph scaling operator and perform eigen-decomposition on the $\mathbf{S_W}$ to obtain the scaling GFT matrix $\mathbf{V}^{-1}_\sigma$\cite{34,35,36}. Further perform eigen-decomposition on $\mathbf{V}^{-1}_\sigma$, and the result can be expressed as
\begin{equation}
    \mathbf{V}^{-1}_\sigma = \mathbf{P_\sigma}\mathbf{J_\sigma}\mathbf{P}^{-1}_\sigma.
\end{equation}
Thus, the weighted adjacency-based GST (wAdj-GST) is defined as
\begin{equation}
    \mathcal{ST}^\sigma_{\mathbf{W}}  = \mathbf{P}_\sigma\mathbf{P}^{-1}_\mathbf{W}.
\end{equation}
\subsubsection{GST based on the Laplacian Matrix}
Unlike the scaling stage described above, \cite{37} constructs a new scaling matrix as
\begin{equation}
    \mathbf{S_L} = \operatorname{diag}\!\left([\sigma^{-\varepsilon r_{0}},\,\sigma^{-\varepsilon r_{1}},\,\ldots,\,\sigma^{-\varepsilon r_{N-1}}]\right),
\end{equation}
where $r_l$ ($l = 0,\dots,N-1$) represents the eigenvalues the eigenvalues corresponding to the eigenvector matrix $\mathbf{P_L}$ and $\varepsilon$ represents the rate of scale change. The dilated graph eigenspace is defined as $\mathbf{S_L}\mathbf{P_L}$. Accordingly, the Laplacian-based GST (Lap-GST) is defined as
\begin{equation}
    \mathcal{ST}^\sigma_{\mathbf{L}} = \mathbf{S_L}\mathbf{P_L}\mathbf{P}^{-1}_\mathbf{L} = \mathbf{S_L}.
\end{equation}
\subsection{GCM}
GCM constitutes a fundamental modulation technique in graph signal processing, enabling precise control over spectral components through linear phase modulation. Currently, two primary GCM implementation approaches have been developed: one based on the weighted adjacency matrix and another utilizing the graph Laplacian matrix.
\subsubsection{GCM based on the Weighted Adjacency Matrix}
There are few theories to define the chirp graph signals. The GCM is defined as an operation applied to the GFT matrix $\mathbf{U}^{-1}_\mathbf{W}$ expressed as:
\begin{equation}
    \mathcal{F}^{\xi}_\mathbf{W} = \mathbf{P_W}\mathbf{J}^{\xi}_\mathbf{W}\mathbf{P}^{-1}_\mathbf{W},
\end{equation}
where $\mathbf{J}^{\xi}_{\mathbf{W}}$ is the matrix of the weighted adjacency-based GCM (wAdj-GCM)\cite{34,35,36} as followed:
\begin{equation}
    \mathcal{CM}^\xi_{\mathbf{W}} = \mathbf{J}^{\xi}_{\mathbf{W}}.
    \label{equ:CM}
\end{equation}
\subsubsection{GCM based on the Laplacian Matrix}
Because defining a chirp signal directly in the vertex domain of a graph is challenging, Chen et al. proposed an alternative approach whereby the frequency of a chirp signal is defined in the spectral domain.
\begin{align}
    \hat{s}_\xi(k) = e^{-\mathbf{i}\frac{\lambda^2_k}{\xi}},
\end{align}
where $\lambda_k$ is the eigenvalues of the Laplacian matrix. Next, the inverse GFT is performed to obtain its vertex domain form as $\mathbf{U_L}\hat{s}_\xi(k)$\cite{37}. Based on this, the Laplician-based GCM (Lap-GCM) is defined as
\begin{equation}
    \mathcal{CM}^\xi_\mathbf{L} = \operatorname{diag}(\mathbf{U_L}\hat{s}_\xi(k)).
    \label{equ1:CM}
\end{equation}
When $\xi = 0$, the frequency of a chirp signal coincides with the frequency of the original signal; applying the IGFT therefore recovers the original signal. In other words, the GCM transform matrix at $\xi = 0$ reduces to the identity matrix $\mathcal{CM}_\mathbf{L}^0 = \mathbf{I}$.
\subsection{GLCT}
\subsubsection{CDDHFs-GLCT}
The LCT is determined by a $2 \times 2$ matrix $\mathbf{M} = (a,b;c,d), a,b,c,d\in\mathbb{R}{}$ with $ad - bc = 1$, where $\mathbf{M}$ is decomposed into FRFT, ST and CM.
\begin{equation}
     \begin{bmatrix}
a & b\\
c & d
\end{bmatrix}
=
\begin{bmatrix}
1 & 0\\
\xi & 1
\end{bmatrix}
\begin{bmatrix}
\sigma & 0\\
0 & \sigma^{-1}
\end{bmatrix}
\begin{bmatrix}
\cos\!{\alpha\pi}/{2} & \sin\!{\alpha\pi}/{2}\\
-\sin\!{\alpha\pi}/{2} & \cos\!{\alpha\pi}/{2}
\end{bmatrix},
\end{equation}
where $\xi$ is the CM parameter, $\sigma$ denotes the ST parameter, and $\alpha$ is the FRFT parameter. There is a one-to-one mapping between $(a,b;c,d)$ and $(\xi, \sigma, \alpha)$: $\xi = \frac{ac+bd}{a^{2}+b^{2}}, \sigma = \sqrt{a^{2}+b^{2}}, \alpha = \frac{2}{\pi}\cos^{-1}\!\left(\frac{a}{\sigma}\right)
      = \frac{2}{\pi}\sin^{-1}\!\left(\frac{b}{\sigma}\right).$

Based on the LCT decomposition framework mentioned above, the GLCT is defined on the basis of different GSOs: the one is a weighted adjacency-based CDDHFs-GLCT (wAdj-CDDHFs-GLCT)\cite{34,35,36}, and the other is a Laplacian-based CDDHFs-GLCT (Lap-CDDHFs-GLCT)\cite{37}.

\subsubsection*{Definition 1} For the matrix $\mathbf{M} = (a,b;c,d)$, denoted $\xi = \frac{ac+bd}{a^{2}+b^{2}}, \sigma = \sqrt{a^{2}+b^{2}}, \alpha = \frac{2}{\pi}\cos^{-1}\!\left(\frac{a}{\sigma}\right)
      = \frac{2}{\pi}\sin^{-1}\!\left(\frac{b}{\sigma}\right)$. The wAdj-CDDHFs-GLCT of $\mathbf{f}$ can be defined as
\begin{equation}
    \hat{\mathbf{f}}_\mathbf{M}^{\mathrm{I}} = \mathbf{J}^{\xi}_\mathbf{W}\mathbf{P_\sigma}\mathbf{P}^{-1}_{\mathbf{W}}\mathbf{P}_{\mathbf{W}}\mathbf{J}^{\beta}_{\mathbf{W}}\mathbf{P}^{-1}_{\mathbf{W}}\mathbf{f} = \mathbf{J}^{\xi}_\mathbf{W}\mathbf{P_\sigma}\mathbf{J}^{\beta}_{\mathbf{W}}\mathbf{P}^{-1}_{\mathbf{W}}\mathbf{f}.
\end{equation}

Analogous to the wAdj-CDDHFs-GLCT, the Lap-CDDHFs-GLCT leverages the spectral properties to define a generalized linear canonical transform  for graph signals.
\subsubsection*{Definition 2}The Lap-CDDHFs-GLCT of $\mathbf{f}$ is defined as
\begin{equation}
\hat{\mathbf{f}}_\mathbf{M}^{\mathrm{II}} = \mathcal{CM}^\xi_\mathbf{L}\mathbf{S_L}\mathbf{P}_\mathbf{L}\mathbf{J}^{\beta}_\mathbf{L}\mathbf{P}^{-1}_\mathbf{L}\mathbf{f}.
\end{equation}

\subsubsection{CM-CC-CM-GLCT}
Compared with the CDDHFs-GLCT in the previous part, the CM-CC-CM-GLCT adopts GCM, GCC and GCM decomposition.
\begin{equation}
 \begin{bmatrix}
a & b\\
c & d
\end{bmatrix} =\begin{bmatrix}
1 & 0\\[2pt]
\xi_1 & 1
\end{bmatrix}
\begin{bmatrix}
1 & b\\[2pt]
0 & 1
\end{bmatrix}
\begin{bmatrix}
1 & 0\\[2pt]
\xi_3 & 1
\end{bmatrix},
\end{equation}
where the first matrix corresponds to GCM with chirp rate $\xi_1 = \frac{d-1}{b}$, and the third matrix has chirp rate $\xi_3 = \frac{a-1}{b}$. The GCC can be futher decomposed as followed,
\begin{equation}
    \begin{bmatrix}
1 & b\\[2pt]
0 & 1
\end{bmatrix} =
\begin{bmatrix}
    0 & -1\\[2pt]
    1 & 0
\end{bmatrix}
\begin{bmatrix}
    1 & 0\\[2pt]
    \xi_2 & 1
\end{bmatrix}
\begin{bmatrix}
    0 & 1\\[2pt]
    -1 & 0
\end{bmatrix}.
\end{equation}
Then
\begin{equation}
\mathbf{M} =\begin{bmatrix}
1 & 0\\[2pt]
\xi_1 & 1
\end{bmatrix}
\begin{bmatrix}
    0 & -1\\[2pt]
    1 & 0
\end{bmatrix}
\begin{bmatrix}
    1 & 0\\[2pt]
    \xi_2 & 1
\end{bmatrix}
\begin{bmatrix}
    0 & 1\\[2pt]
    -1 & 0
\end{bmatrix}
\begin{bmatrix}
1 & 0\\[2pt]
\xi_3 & 1
\end{bmatrix},
\label{equ:cm-cc-cm-glct}
\end{equation}
where the second GCM matrix has chirp rate $\xi_2 = -b$. We obtain CM-CC-CM-GLCT composed of GFT, IGFT, and GCM\cite{35}. 

\subsubsection*{Definition 3}For the matrix $\mathbf{M} = (a,b;c,d)$, denoted $\xi_1 = \frac{d-1}{b}, \xi_2 = -b, \xi_3 = \frac{a-1}{b}$, the wAdj-CM-CC-CM-GLCT of $\mathbf{f}$ is defined as 
\begin{equation}
    \hat{\mathbf{f}}_\mathbf{M}^{\mathrm{III}} = \mathbf{J}^{\xi_1}_\mathbf{W}\mathbf{U}_\mathbf{W}\mathbf{J}^{\xi_2}_\mathbf{W}\mathbf{U}^{-1}_\mathbf{W}\mathbf{J}^{\xi_3}_\mathbf{W}\mathbf{f}.
\end{equation}
\section{Lap-CM-CC-CM-GLCT}
\subsection{Definition}
To achieve a graph domain implementation with desirable additivity and invertibility within a unified LCT framework, this section adopts a Laplacian spectral basis expansion. We propose a new CM-CC-CM-GLCT in the eigenbasis of the Laplacian matrix (Lap-CM-CC-CM-GLCT). Similar to the wAdj-CM-CC-CM-GLCT, the Lap-CM-CC-CM-GLCT likewise adopts GCM, GCC, GCM decomposition and the GCC can futher be decomposed into the form of GFT, GCM and IGFT like \eqref{equ:cm-cc-cm-glct}. 
\subsubsection*{Definition 4}The Lap-CM-CC-CM-GLCT of $\mathbf{f}$ is defined as
\begin{equation}
  \hat{\mathbf{f}}_\mathbf{M}^{\mathrm{IV}} = \mathcal{F}_\mathbf{L}^\mathbf{M}\mathbf{f} =\mathcal{CM}_{\mathbf{L}}^{\xi_1}\mathbf{U}_\mathbf{L}\mathcal{CM}_{\mathbf{L}}^{\xi_2}
\mathbf{U}^{-1}_\mathbf{L}\mathcal{CM}_{\mathbf{L}}^{\xi_3}\mathbf{f},
\end{equation}
where $\xi_1 = \frac{d-1}{b}$, $\xi_2 = -b$, $\xi_3 = \frac{a-1}{b}$.

\subsection{Properties}
In this part, we introduce some properties of the Lap-CM-CC-CM-GLCT.
\subsubsection*{Property 1 (Linearity)}The Lap-CM-CC-CM-GLCT is a linear transformation:
\begin{equation}
    \mathcal{F}_\mathbf{L}^\mathbf{M}\{\alpha\mathbf{f}+\beta \mathbf{g}\} = \alpha \mathcal{F}_\mathbf{L}^\mathbf{M}\{\mathbf{f}\} 
    + \beta \mathcal{F}_\mathbf{L}^\mathbf{M}\{\mathbf{g}\}.
\end{equation}
\subsubsection*{Property 2 (Zero rotation)}

When $\mathbf{M} = \mathbf{I}$, i.e., $\xi_1 = \xi_2 = \xi_3 = 0$, the Lap-CM-CC-CM-GLCT reduce to identity. 
\begin{equation}
\mathcal{F}_\mathbf{L}^\mathbf{I} = \mathbf{I}.
\end{equation}
\subsubsection*{Property 3 (Additivity)}Let $\mathbf{M_1}, \mathbf{M_2} \in\mathrm{SL}(2,\mathbb{R})$, for the Lap-CM-CC-CM-GLCT, the additivity property holds 
\begin{equation}
\mathcal{F}^{\mathbf{M_2}}_\mathbf{L}\mathcal{F}^{\mathbf{M_1}}_\mathbf{L} = \mathcal{F}^{\mathbf{M_2M_1}}_\mathbf{L}.
\end{equation}
It implies that the cascade of
several GLCTs with parameter matrices can be replaced by only one GLCT using parameter matrix.
\subsubsection*{Property 4 (Invertibility)}For any matrix $\mathbf{M}\in\mathrm{SL}(2,\mathbb{R})$,
\begin{equation}    (\mathcal{F}^{\mathbf{M}}_\mathbf{L})^{-1} = \mathcal{F}^{\mathbf{M}^{-1}}_\mathbf{L}.
\end{equation}

\subsubsection*{Property 5 (Unitarity)}The Lap-CM-CC-CM-GLCT matrix is a unitary matrix, i.e., it satisfies
\begin{equation}
    \mathcal{F}_\mathbf{L}^\mathbf{M}(\mathcal{F}_\mathbf{L}^\mathbf{M})^\mathrm{H}
= \mathbf{I}.
\end{equation}
\subsubsection*{Proof}The proofs of Properties 1 and 2 are straightforward and thus omitted. See Appendix A for the proofs of Properties 3--5.\qed
\section{GWF in the Transform Spectral Domain}
\subsection{GLCT Wiener Filtering Framework}
Consider the graph observation model:
\begin{equation}
    \mathbf{\tilde{f}} = \mathbf{Gf + n},
\end{equation}
where $\mathbf{G}$ is a known perturbation matrix, and $\mathbf{f}$ is a smooth signal. Moreover, $\mathbf{n}$ is the additive noise term. 

Before presenting the new problem, we begin with Wiener filtering in the graph domain. Let $\mathbf{f}$ is denotes the original graph signal and $\mathbf{\tilde{f}}$ the corresponding observation. The optimal Wiener filtering problem\cite{46,47} can be formalized as:
\begin{equation}
    \min_{\mathbf{R}}\; \mathbb{E}\left\{ \left\| \mathbf{R}\,\mathbf{\tilde{f}} - \mathbf{f} \right\|_2^{2} \right\}.
\end{equation}

This paper aims to develop an optimal filtering approach in the transform spectral domain to recover the original signal $\mathbf{f}$ with the minimal MSE. Inspired by Wiener filtering in the GFRFT domain, we extend this concept to the GLCT domain under various definitions.  Accordingly, the estimated signal can be expressed as $\mathcal{F}^{\mathbf{M}^{-1}}\mathbf{H}\mathcal{F}^{\mathbf{M}}\mathbf{\tilde{f}}$, where $\mathcal{F}^{\mathbf{M}}$ denotes the GLCT and its superscript $\mathbf{M}$ represents the parameter matrix of this transform. In this framework, the observed signal $\mathbf{\tilde{f}}$ is processed by a forward GLCT, filtered in the transform spectral domain, and then inverted back to the vertex domain, which can be stated as:
\begin{equation}
    \mathrm{MSE} = \mathbb{E}\left\{ \left\|\mathcal{F}^{\mathbf{M}^{-1}}\mathbf{H}\mathcal{F}^{\mathbf{M}}\mathbf{\tilde{f}} - \mathbf{f} \right\|_2^{2} \right\}.
\end{equation}
Therefore, the objective is to design the matrix $
\mathbf{H}$ that solves the following minimization problem\cite{46,47,48}:
\begin{equation}
   \mathbf{H}^{*} =  \arg\min_{\mathbf{H}}\; \mathbb{E}\left\{ \left\|\mathcal{F}^{\mathbf{M}^{-1}}\mathbf{H}\mathcal{F}^{\mathbf{M}}\mathbf{\tilde{f}} - \mathbf{f} \right\|_2^{2} \right\}.
    \label{equ:filter}
\end{equation}

In discrete signal processing, a multiplicative filter is modeled as a diagonal matrix. We thus force $\mathbf{H}$ in \eqref{equ:filter} to be diagonal. This parameterization reduces the problem to finding an optimal non-zero vector $\mathbf{h} = [h_1, h_2, \ldots, h_N]^\mathrm{T}$ in $\mathbb{C}^N$.  Let $\mathcal{F}^{\mathbf{M}^{-1}} = [\mathbf{w}_1, \mathbf{w}_2,\ldots,\mathbf{w}_N]$, $\mathcal{F}^{\mathbf{M}} = [\mathbf{\tilde{w}}_1, \mathbf{\tilde{w}}_2,\ldots,\mathbf{\tilde{w}}_N]^\mathrm{T}$, and $\mathbf{W}_i = \mathbf{w}_i\tilde{\mathbf{w}}_i$. Introducing the matrices $\mathbf{W}_i$, the problem simplifies to designing the vector $\mathbf{h}$ that solves the following minimization:
\begin{equation}
\mathbf{h}^{*}=
\arg\min_{\mathbf{h}\in\mathbb{C}^{N}}
\; \mathbb{E}\!\left\{
\left\| \sum_{i=1}^{N} h_i\,\mathbf{W}_i\,\mathbf{\tilde{f}} - \mathbf{f} \right\|_2^{2}
\right\}.
\end{equation}
 The optimal filter coefficients $\mathbf{h}^{*}$ are obtained by solving the Wiener-Hopf equation: $\mathbf{Th = q}$, as described in \cite{46}. Provided that $\mathbf{T}$ is invertible, the optimal filter cofficients $\mathbf{h}^{*} = \mathbf{T}^{-1}\mathbf{q}$.  In our optimization strategy, the free parameters $a$, $b$, and $d$ of the GLCT are selected through an exhaustive grid search, with the objective of minimizing the MSE of the signal after filtering.
\begin{equation}
    (a^{*}, b^{*}, d^{*}) = \arg\min_{a,b,d}\mathbb{E}\left\{ \left\|\mathcal{F}^{\mathbf{M}^{-1}}\mathbf{H}^{*}\mathcal{F}^{\mathbf{M}}\mathbf{\tilde{f}} - \mathbf{f} \right\|_2^{2} \right\}.
\end{equation}

In general, the optimal filter is derived in a two-step process. Specifically, the Wiener-Hopf equation is solved for any fixed parameters $(a, b, d)$, yielding an optimal diagonal filter $\mathbf{H}^{*}$ with coefficients that are functions of these parameters. The overall optimum is then found by performing a grid search over $(a, b, d)$ to minimize the resulting MSE. The complete
procedure is presented as pseudocode in Algorithm 1. These settings are used for subsequent experiments.
\begin{algorithm}[t]
\caption{Grid search for GLCT parameters and Wiener filter}
\label{alg:grid-glct}
\KwIn{

Input graph signal: 
$\mathbf{f}$; Target signal: $\mathbf{\tilde{f}}$.\\
Parameter grids: $\mathcal{A}$ (for $a$), $\mathcal{B}$ (for $b$), $\mathcal{D}$ (for $d$);\\
GLCT basis choice ($\mathbf{U}_\mathbf{W}$ or $\mathbf{U}_\mathbf{L}$).
}
\KwOut{Best parameters $(a^\ast,b^\ast,d^\ast)$; best filter $\mathbf{H}^\ast$; best loss $\mathrm{MSE}^\ast$.}

\BlankLine
\textbf{1:} Precompute spectral basis (eigen-decomposition).\\
\textbf{2:} Compute the loss.

\For{$a\in\mathcal{A}$}{
  \For{$b\in\mathcal{B}$}{
    \For{$d\in\mathcal{D}$}{
      \textbf{3:} Construct $a,b,d$ and the GLCT operators
      $\mathcal{F}^{\mathbf{M}}$ and $\mathcal{F}^{\mathbf{M}^{-1}}$.\\
      \textbf{4:} Solve Wiener-Hopf for this candidate:
      $\mathbf{h} = \mathbf{T}^{-1}\mathbf{q}$,
      where $\mathbf{H}=\operatorname{diag}(\mathbf h)$.\\
      \textbf{5:} Evaluate loss (MSE):
      $\mathrm{MSE}(\mathbf{H},a,b,d)$.\\
      \textbf{6:} If $\mathrm{MSE}^\ast < \mathrm{MSE}$\\
      {$\mathrm{MSE}^\ast \leftarrow \mathrm{MSE}$;
      $(a^\ast,b^\ast,d^\ast) \leftarrow (a,b,d)$;
      $\mathbf{H}^\ast \leftarrow \mathbf{H}$;}
    }
  }
}
\textbf{7:} \Return $(a^\ast,b^\ast,d^\ast)$, $\mathbf{H}^\ast$, $\mathrm{MSE}^\ast$.
\end{algorithm}

\subsection{Joint Trainable Transform-Fliter Optimization}
 Given that grid search-based parameter selection struggles to scale efficiently in high-dimensional parameter spaces, this paper proposes an end-to-end differentiable framework that formulates optimal Wiener filtering as a joint optimization problem over the free parameters of the GLCT and the transform spectral domain diagonal filter. Intuitively, instead of enumerating candidate combinations of $(a, b, d)$, we treat them along with the filter coefficients—as learnable variables, and directly minimize the reconstruction error on training data via gradient descent.
\subsubsection*{Theorem 1}The loss function $\mathrm{MSE}$ of the proposed GLCT-GWF framework is differentiable with respect to the transform parameters $a,b,d$ and the filtering coefficients $\mathbf{H}$. The gradients $\nabla_{\mathbf {H}}\mathrm{MSE}$, $\nabla_{a}\mathrm{MSE}$, $\nabla_{b}\mathrm{MSE}$, $\nabla_{d}\mathrm{MSE}$ exist and can be computed via backpropagation.
\subsubsection*{Proof}See Appendix B.\qed

Based on Theorem 1, the core idea of this method is to break through the limitations of traditional step-by-step optimization by constructing the entire signal processing pipeline into a complete and differentiable computational graph. Within this framework, the free parameters of the GLCT are no longer predefined enumerated candidate values but instead become trainable variables that can be iteratively updated alongside the filter coefficients. By adopting MSE as the target loss function and employing the backpropagation algorithm to automatically compute its gradient with respect to all parameters, the gradient descent method can simultaneously and synergistically adjust the shape of the transform domain and the characteristics of the filter. The complete
procedure is presented as pseudocode in Algorithm 2. These settings are used for subsequent experiments.

\begin{algorithm}[t]
\caption{Adam-based joint transform-filter optimization algorithm}
\label{alg:sgd-filter}
\KwIn{

Input graph signal: 
$\mathbf{f}$; Target signal: $\mathbf{\tilde{f}}$.\\
Initial filter: $\mathbf{H}^{0}=\operatorname{diag}(\mathbf h^{0})$.\\

Initial GLCT parameters: $(a^{0},b^{0},d^{0})$.\\
Learning rate: $\varepsilon$; \\
Stopping criterion; Maximum iterations.
}
\KwOut{

Learned filter: $\mathbf{H}^{\mathrm{adam}}=\operatorname{diag}(\mathbf h^{\mathrm{adam}})$.\\
Learned GLCT parameters: 
$(a^{\mathrm{adam}},b^{\mathrm{adam}},d^{\mathrm{adam}})$
Loss value: $\mathrm{MSE}(\mathbf{H}^{\mathrm{adam}}, a^{\mathrm{adam}}, b^{\mathrm{adam}}, d^{\mathrm{adam}})$.
}

\textbf{1:} Compute GLCT transform operator $\mathcal{F}_{\mathbf X}^{\mathbf{M}}$.\\
\textbf{2:} Evaluate the initial loss $\mathrm{MSE}(\mathbf{H}^{0}, a^{0}, b^{0},d^{0})$.\\
$
\begin{aligned}
\textbf{3: }\text{Set } & 
\mathbf{H}^{\mathrm{adam}} \leftarrow \mathbf{H}^{0} \\
& a^{\mathrm{adam}} \leftarrow a^{0},\quad
b^{\mathrm{adam}} \leftarrow b^{0},\quad
d^{\mathrm{adam}} \leftarrow d^{0}.
\end{aligned}
$

\textbf{4:} \While{stopping criterion not met}{
  Compute the gradient of the loss: 
  \[
  \nabla_{\mathbf H^{\mathrm{adam}}}\,\mathrm{MSE}, \nabla_{a^{\mathrm{adam}}}\mathrm{MSE},\nabla_{b^{\mathrm{adam}}}\mathrm{MSE},\nabla_{d^{\mathrm{adam}}}\mathrm{MSE}.
  \] 
  {Filter update:}\quad
  $\mathbf{H}^{\mathrm{adam}} \leftarrow
   \mathbf{H}^{\mathrm{adam}} - \varepsilon\,\nabla_{\mathbf {H}^{\mathrm{adam}}}\mathrm{MSE}$.\\
  {Parameters update:}\quad
  $a^{\mathrm{adam}} \leftarrow
   a^{\mathrm{adam}} - \varepsilon\,\nabla_{a^{\mathrm{adam}}}\mathrm{MSE}$,\quad
   $b^{\mathrm{adam}} \leftarrow
   b^{\mathrm{adam}} - \varepsilon\,\nabla_{b^{\mathrm{adam}}}\mathrm{MSE}$,\quad
   $d^{\mathrm{adam}} \leftarrow
   d^{\mathrm{adam}} - \varepsilon\,\nabla_{d^{\mathrm{adam}}}\mathrm{MSE}$.
}
\textbf{5:} {Compute the final loss} $\mathrm{MSE}(\mathbf{H}^{\mathrm{adam}}, a^\mathrm{adam}, b^\mathrm{adam}, d^\mathrm{adam})$.
\end{algorithm}
\subsection{Computational Complexity Analysis}
To enable a fair comparison, we evaluate the computational of the four GLCT variants under a unified setting: the graph has size $N$; dense matrix-matrix products cost $\mathbf{O}(N^3)$; the computational cost of diagonal multiplication is $\mathbf{O}(N^2)$ which can be negligible. For a single forward transform, Lap-CDDHFs-GLCT involves only one dense matrix multiplication and a few diagonal multiplications, yielding a time complexity of $\mathbf{O}(N^3)$. In contrast, wAdj-CDDHFs-GLCT, wAdj-CDDHFs-GLCT, and Lap-CM-CC-CM-GLCT each require two dense matrix multiplications, resulting in approximately $2\mathbf{O}(N^3)$. The inverse transform is of the same order. Detailed quantitative comparisons of their computational overhead are summarized in \Cref{tab:comparison}.
\begin{table}[h]
\centering
\caption{Comparison Between Proposed Method and Existing Approachs}
\label{tab:comparison}
\setlength{\tabcolsep}{0.6pt}
\begin{tabular}{c c c c}
\hline
\textbf{Method} & \makecell{\textbf{Dense Matrix} \\ \textbf{Multiplications}} & \makecell{\textbf{Diagonal Matrix} \\ \textbf{Multiplications}} & \makecell{\textbf{Computational} \\ \textbf{Complexity}} \\
\hline
wAdj-CDDHFs-GLCT & $2$ & $2$ & $2\mathbf{O}(N^3)$ \\
wAdj-CM-CC-CM-GLCT & $1$ & $3$ & $\mathbf{O}(N^3)$ \\
Lap-CDDHFs-GLCT & $2$ & $3$ & $2\mathbf{O}(N^3)$ \\
Lap-CM-CC-CM-GLCT & $2$ & $3$ & $2\mathbf{O}(N^3)$ \\
\hline

\end{tabular}
\end{table}
All four methods share a one-time spectral-basis construction cost (e.g., eigen-decomposition of the Laplacian or weighted adjacency matrix), which is $\mathbf{O}(N^3)$ and can be amortized across runs.

For each set of free parameters, the optimal filter coefficients are obtained by solving the Wiener-Hopf equation $\mathbf{Th = q}$. Constructing $\mathbf{T}$ cost $\mathbf{O}(N^4)$, constructing $\mathbf{q}$ cost $\mathbf{O}(N^3)$, and solving the linear system(e.g., via Cholesky) cost $\mathbf{O}(N^3)$. Hence, the per parameter setting is dominated by forming $\mathbf{T}$ and equals $\mathbf{O}(N^4)$\cite{46}. Let the GLCT have three free parameters $(a, b, d)$. Suppose the grid contains $n_a, n_b, n_d$ samples along each axis. Since one evaluation at a fixed $(a, b, d)$ requires forming $\mathbf{T}$ and $\mathbf{q}$ and solving $\mathbf{Th = q}$, with cost dominated by building $\mathbf{T}$ at $\mathbf{O}(N^4)$, the overall complexity of grid search is $\mathbf{O}(n_an_bn_dN^4)$.

The core computational steps of the Adam-based joint transform-filter learning algorithm include the generation of the GLCT operator, the calculation of the loss function and its gradients and the parameter updates. Each forward pass involves sequentially applying the GLCT, spectral filtering with complexity $\mathbf{O}(N)$, inverse GLCT, and MSE calculation, resulting in a forward computational complexity per iteration of $\mathbf{O}(N^2)$. During backpropagation, gradients are computed via automatic differentiation, whose complexity is typically a constant multiple of the forward pass, thus also $\mathbf{O}(N^2)$. Parameter updates are performed using the Adam optimizer, with a per-parameter update complexity of $\mathbf{O}(1)$. The parameters to be updated include an $N$-dimensional filter coefficient vector and three scalar GLCT parameters, leading to a total parameter update complexity of $\mathbf{O}(N)$ per iteration. Assuming a maximum of $K$ iterations, the overall computational complexity of the algorithm is $\mathbf{O}(KN^2)$. These computational complexity results for both methods are summarized in \Cref{tab:comparison 1}. 
\begin{table}[h]
\centering
\caption{Comparison of Computational Complexity}
\label{tab:comparison 1}
\setlength{\tabcolsep}{7.0pt}
\begin{tabular}{c c c c}
\hline
\textbf{Method}  & \makecell{\textbf{Overall Computational } \\ \textbf{Complexity}} \\
\hline
Grid Search-based Method  & $\mathbf{O}(n_an_bn_dN^4)$ \\
Adam-based Joint Optimization Method  & $\mathbf{O}(KN^2)$ \\
\hline
\end{tabular}
\end{table}

In general, all four GLCT share a one-time spectral-basis construction cost of $\mathbf{O}(N^3)$. As the computational complexity comparison shows, the grid search-based method requires executing an operation with complexity of $\mathbf{O}(N^4)$ for constructing a matrix over the entire three-dimensional parameter grid, resulting in an overall complexity that grows cubically with the number of sampling points and incurs extremely high computational cost. In contrast, the Adam-based joint optimization method transforms the computational complexity into a linear function of the number of iterations $K$, with each iteration requiring only $\mathbf{O}(N^2)$ computations, significantly improving computational efficiency. This difference makes the joint optimization method particularly suitable for high-dimensional parameter optimization problems and large-scale GSP tasks.

\subsection{Convexity and Convergence Analysis}Despite the non-convex optimization landscape formed by adjusting the GLCT parameters $a, b, d$ in grid search methods, the subproblem of solving for $\mathbf{h}$ under any fixed parameters
is strictly convex\cite{46,48}. In practical applications, once candidate parameters is given, the filter coefficients $\mathbf{h}$ can be obtained as a unique global optimal solution. To address the inherent non-convexity of the problem, we discretize the parameters over a grid and solve the convex filter design problem at each grid point. The resulting MSE surface demonstrates smooth behavior.

In contrast, for the Adam-based optimization algorithms, the GLCT incorporates parameters into the forward-diagonal-inverse pipeline in a non-linear manner. The MSE objective function with respect to these parameters is composed of multiple nonlinear operators, typically resulting in non-convexity that may contain multiple local minima and saddle points. Consequently, joint optimization of the parameters and $\mathbf{h}$ using gradient descent is inherently non-convex.

When the parameters are fixed, the optimal filtering problem is convex and admits a closed-form solution. With grid search, refining the grid can approach the continuous domain optimum, but the computational cost grows rapidly with the number of candidates. In contrast, the Adam-based joint optimization eliminates the need for candidate enumeration, incurs lower computational burden, and achieves stable objective descent with convergence to first-order critical points under proper regularization and step size settings, albeit the problem is intrinsically non-convex. Given the distinct focuses of these two approaches, this paper does not simply compare their experimental performances but selects the appropriate optimization strategy based on practical application scenarios: grid search is preferred when the search space is small or strong priors exist with an emphasis on global interpretability; joint Adam-based methods are adopted when the parameter space is large and data-adaptive optimization is required.

\begin{figure*}
        \centering
    \includegraphics[width=0.6\linewidth]{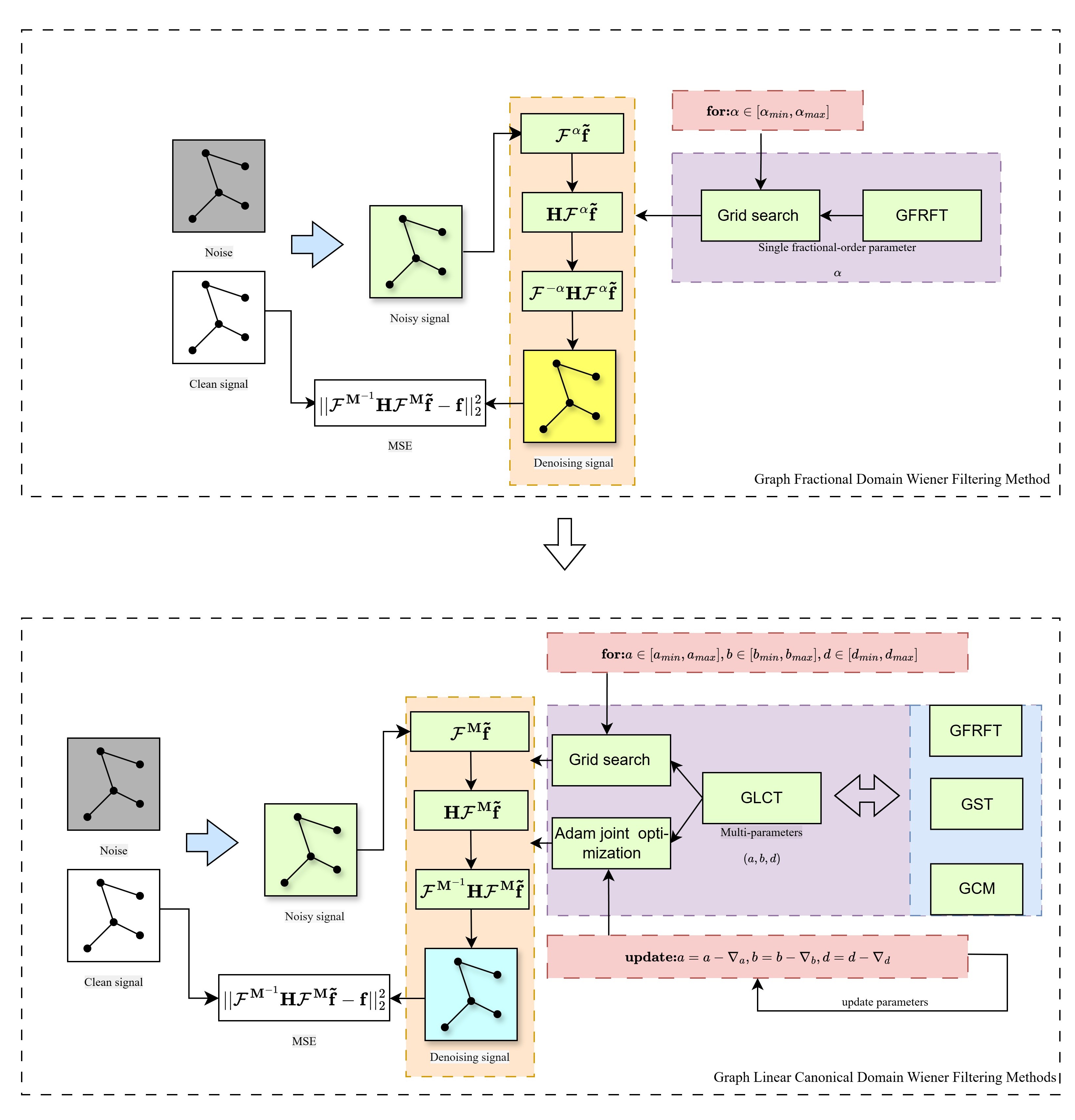}
    \caption{Flowchart of GLCT based transform domain filtering: comparison between traditional single parameter grid-search strategy (Top) and the proposed multi-parameter joint optimization framework (Bottom).}
    \label{fig:GLCT流程图}
\end{figure*}

To visually illustrate the aforementioned method and the evolution from traditional denoising strategies to the proposed framework, a flowchart of the filtering task is presented in Fig. \ref{fig:GLCT流程图}.

\section{Numerical Experiments \& Results}
This section presents experiments conducted on both synthetic and real-world datasets to validate the performance of existing and proposed GLCTs denoising methods.
\subsection{Synthetic Data}
\begin{figure*}[!t]
    \centering
    \begin{minipage}{0.32\textwidth}
        \centering
        \includegraphics[width=\linewidth]{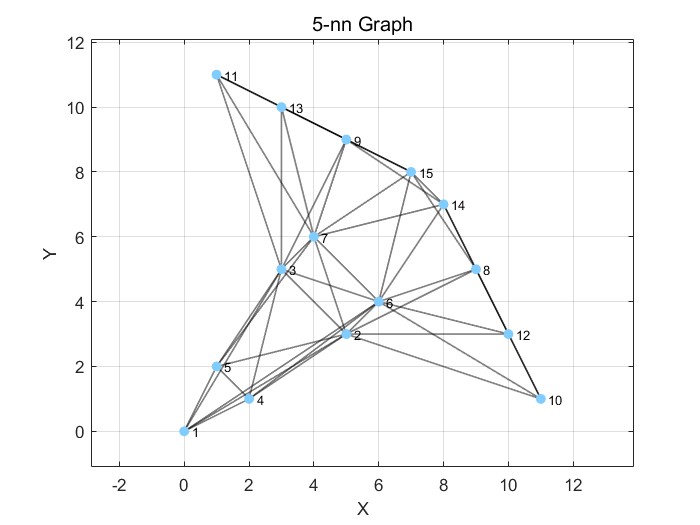}
        \caption{5-nn graph with 15 nodes.}\label{fig:5-nn}
    \end{minipage}\hfill
    \begin{minipage}{0.32\textwidth}
        \centering
        \includegraphics[width=\linewidth]{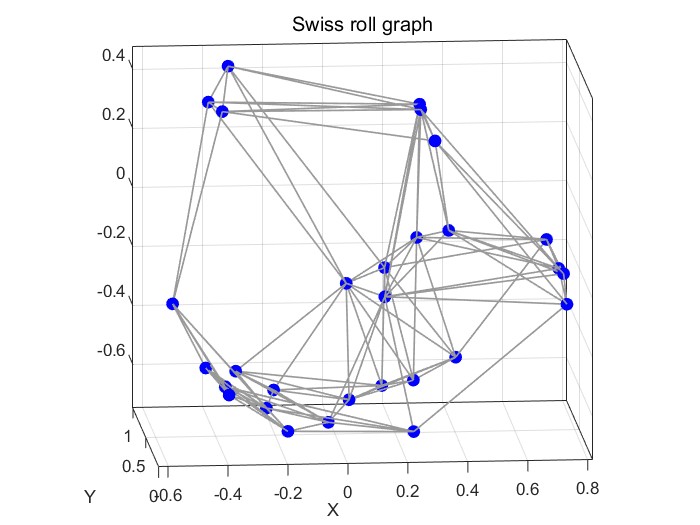}
        \caption{Swiss roll graph with 30 nodes.}\label{fig:Swiss roll}
    \end{minipage}\hfill
    \begin{minipage}{0.32\textwidth}
        \centering
        \includegraphics[width=\linewidth]{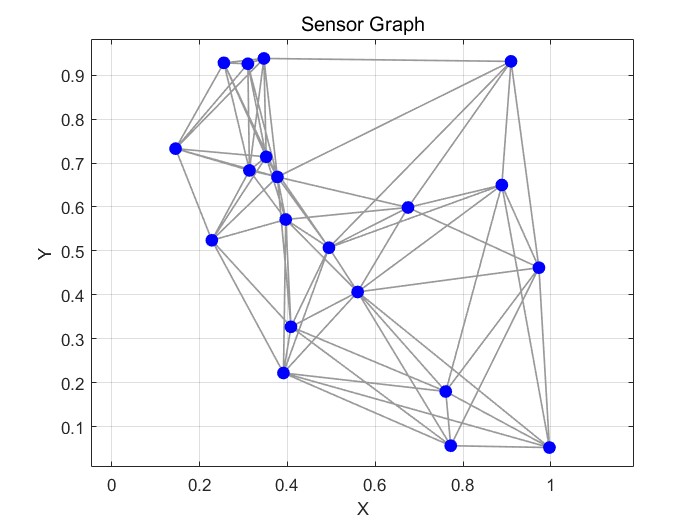}
        \caption{Sensor graph with 20 nodes.}\label{fig:Sensor}
    \end{minipage}
\end{figure*}

We conducted experiments on three types of synthetic graphs: 5-nearest neighbor (5-nn) graph with 15 nodes\cite{46}, Swiss roll graph with 30 nodes, and sensor graph with 20 nodes, as shown in Figs. \ref{fig:5-nn}--\ref{fig:Sensor}. For the 5-nn graph, nodes are randomly distributed in a 2D plane, with each node connected to its 5 nearest neighbors. The graph signal $\mathbf{f}$ is modeled using an autocorrelation matrix: $\mathbb{E}\{\mathbf{x}\mathbf{x}^{\mathrm{H}}\} = \frac{1}{\lambda_{\max}(\mathbf{C})}\mathbf{C}$, where $\lambda_{\max}(\mathbf{C})$ is the largest eigenvalues of $\mathbf{C} \in \mathbb{R}^{N\times N}$ and for any $1 \leq i,j \leq N$\cite{46},
\[
\mathbf{C}_{ij} = 
\begin{cases} 
2, & \text{if } i = j \\ 
1, & \text{if nodes } i \text{ and } j \text{ are connected} \\ 
0, & \text{otherwise}.
\end{cases}
\]
Additive Gaussian noise with zero mean and variance $s^2\mathbf{I}$ for $(s \in \{0.5, 1.0, 1.5\})$ is added to test robustness.

Given the small size of the synthetic graphs used in this section, we adopt grid search over the transform parameters, and solve a Wiener filter once at each grid point. The settings are as follows: 
\begin{itemize}
    \item GFRFT: fractional order $\alpha \in [0,2]$ with a step of 0.1.
    \item GLCT: parameters $a, b, d$ are searched over $[0,2]$ with a step of 0.1. To avoid degeneracy at $b = 0$, we enforce $b \in [0.1, 2]$.
\end{itemize}
\begin{table}[h]
\centering
\caption{Denoising Results (MSE) on 5-nn Graph}
\setlength{\tabcolsep}{10pt}
\begin{tabular}{cccc}
\toprule
\textbf{Method} & $s = 0.5$ & $s = 1.0$ & $s = 1.5$ \\
\midrule
GFRFT$_\mathbf{W}$ & 2.63305 & 2.75315 & 2.88111 \\
GFRFT$_\mathbf{L}$ & 2.84614 & 2.94902 & 3.04439 \\
wAdj-CDDHFs-GLCT & \textbf{2.53739} & \textbf{2.70592} & \textbf{2.86640} \\
wAdj-CM-CC-CM-GLCT & 2.58884 & 2.74056 & 2.88810 \\
Lap-CDDHFs-GLCT & 2.65785 & 2.75978 & 2.87515 \\
Lap-CM-CC-CM-GLCT & 2.59918 & 2.71985 & 2.84557 \\
\bottomrule
\end{tabular}
\label{tab:denoising_results_5nn}
\end{table}

\begin{table}[h]
\centering
\caption{Denoising Results (MSE) on Swiss Roll Graph}
\setlength{\tabcolsep}{10pt}
\begin{tabular}{cccc}
\toprule
\textbf{Method} & $s = 0.5$ & $s = 1.0$ & $s = 1.5$ \\
\midrule
GFRFT$_\mathbf{W}$ & 5.41332 & 5.49585 & 5.59003 \\
GFRFT$_\mathbf{L}$ & 5.29207 & 5.37646 & 5.47938 \\
wAdj-CDDHFs-GLCT & \textbf{5.11582} & \textbf{5.22133} & \textbf{5.34875} \\
wAdj-CM-CC-CM-GLCT & 5.31804 & 5.39537 & 5.49158 \\
Lap-CDDHFs-GLCT & 5.23232 & 5.32352 & 5.43279 \\
Lap-CM-CC-CM-GLCT & 5.25209 & 5.32666 & 5.42166 \\
\bottomrule
\end{tabular}
\label{tab:denoising_results_swiss}
\end{table}

\begin{table}[h]
\centering
\caption{Denoising Results (MSE) on Sensor Graph}
\setlength{\tabcolsep}{10pt}
\begin{tabular}{cccc}
\toprule
\textbf{Method} & $s = 0.5$ & $s = 1.0$ & $s = 1.5$ \\
\midrule
GFRFT$_\mathbf{W}$ & 3.38296 & 3.50985 & 3.63292 \\
GFRFT$_\mathbf{L}$ & 3.43203 & 3.53418 & 3.63919 \\
wAdj-CDDHFs-GLCT & \textbf{3.06556} & \textbf{3.23947} & \textbf{3.41947} \\
wAdj-CM-CC-CM-GLCT & 3.16949 & 3.33281 & 3.49228 \\
Lap-CDDHFs-GLCT & 3.42717 & 3.51943 & 3.62087 \\
Lap-CM-CC-CM-GLCT & 3.22928 & 3.37772 & 3.52021 \\
\bottomrule
\end{tabular}
\label{tab:denoising_results_sensor}
\end{table}

\begin{figure*}[!t]
\centering
\subfloat[]{\includegraphics[width=0.32\textwidth]{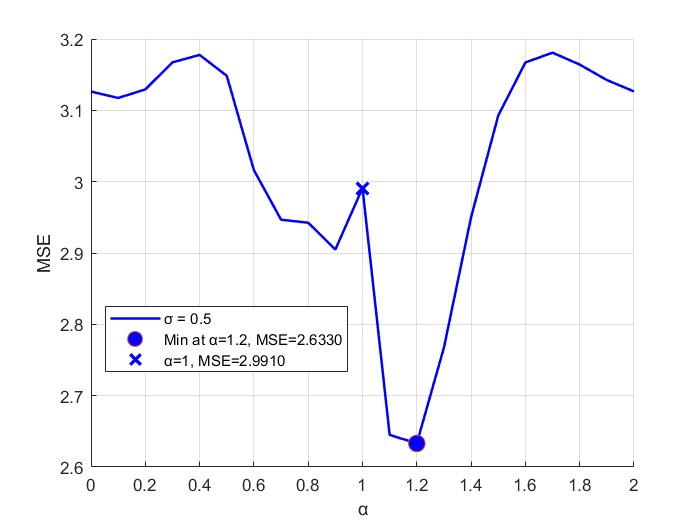}\label{fig:sub1}}\hfill
\subfloat[]{\includegraphics[width=0.32\textwidth]{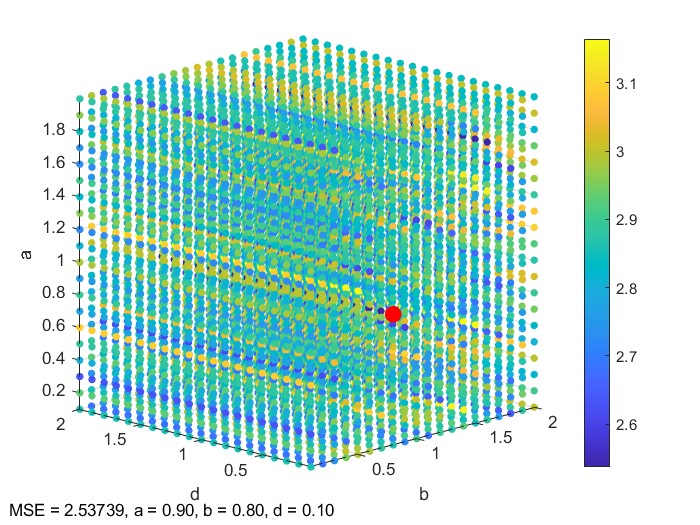}\label{fig:sub2}}\hfill
\subfloat[]{\includegraphics[width=0.32\textwidth]{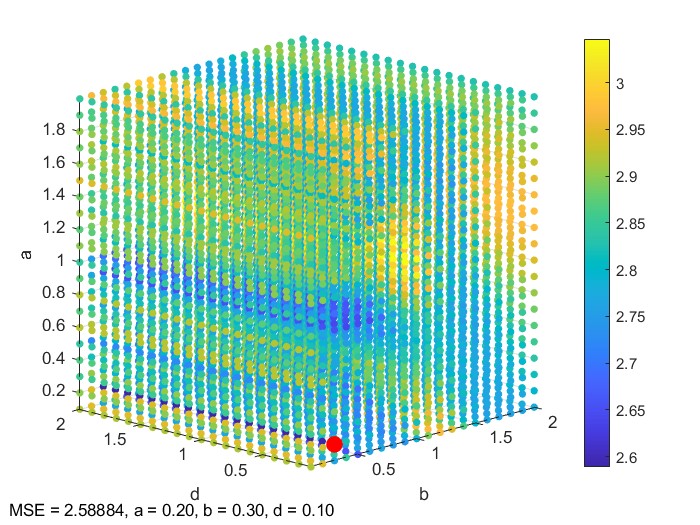}\label{fig:sub3}}\\[0.5em]
\subfloat[]{\includegraphics[width=0.32\textwidth]{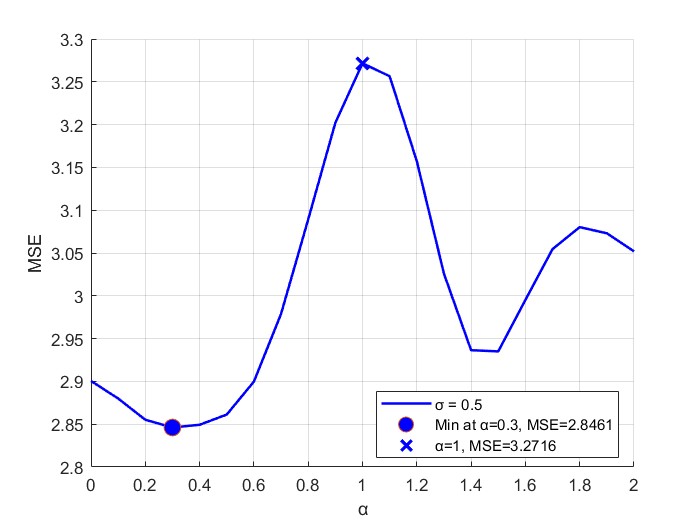}\label{fig:sub4}}\hfill
\subfloat[]{\includegraphics[width=0.32\textwidth]{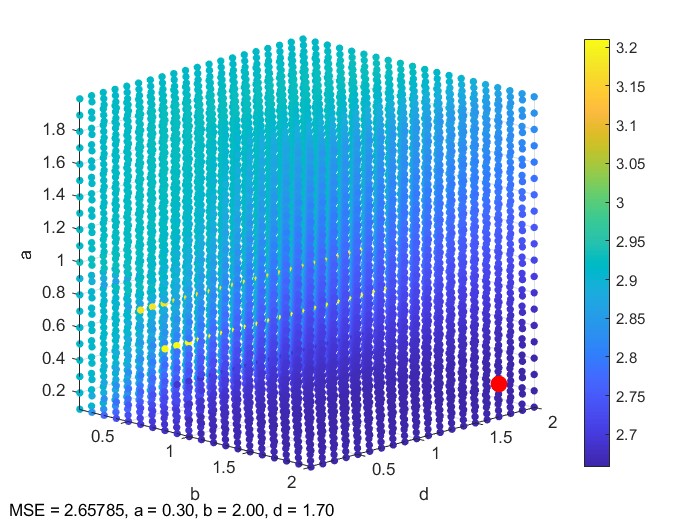}\label{fig:sub5}}\hfill
\subfloat[]{\includegraphics[width=0.32\textwidth]{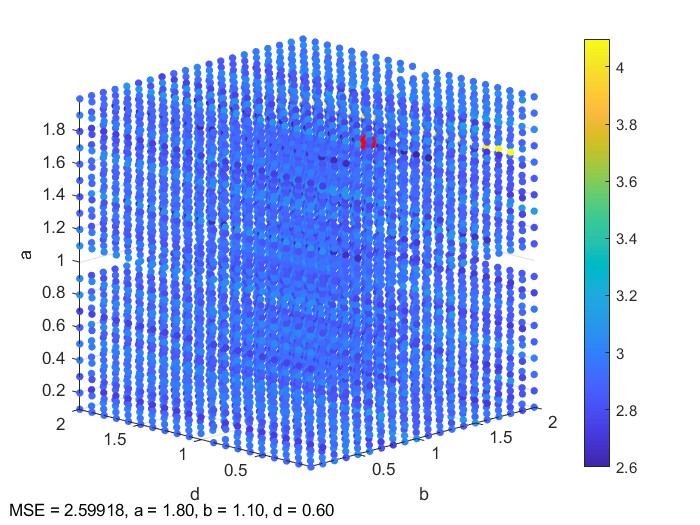}\label{fig:sub6}}
\caption{Denoising results (MSE) on a 5-nn graph ($s = 0.5$) : comparison of GFRFT and GLCT under different GSOs. (a)-(c) use the weighted adjacency matrix; (d)-(f) use the Laplacian matrix; b: wAdj-CDDHFs-GLCT; c: wAdj-CM-CC-CM-GLCT; e: Lap-CDDHFs-GLCT; f: Lap-CM-CC-CM-GLCT.}
\label{fig:5-nn-MSE}
\end{figure*}

\begin{figure*}[!t]
\centering
   \subfloat[]{\includegraphics[width=0.32\textwidth]{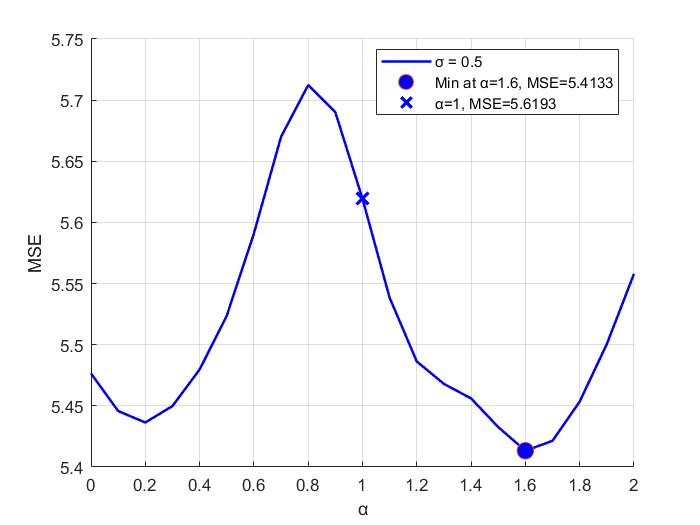}\label{fig:sub7}}\hfill
\subfloat[]{\includegraphics[width=0.32\textwidth]{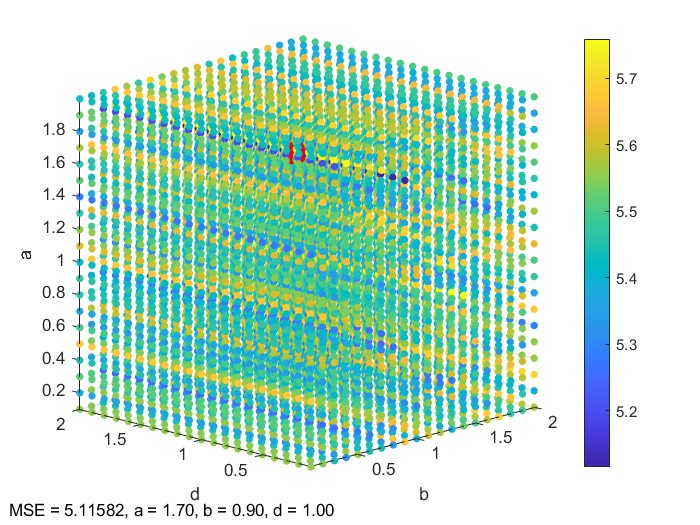}\label{fig:sub8}}\hfill
\subfloat[]{\includegraphics[width=0.32\textwidth]{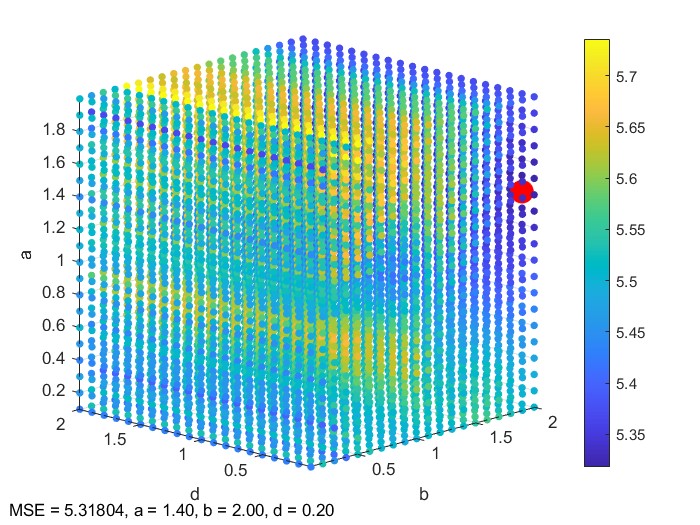}\label{fig:sub9}}\\[0.5em]
\subfloat[]{\includegraphics[width=0.32\textwidth]{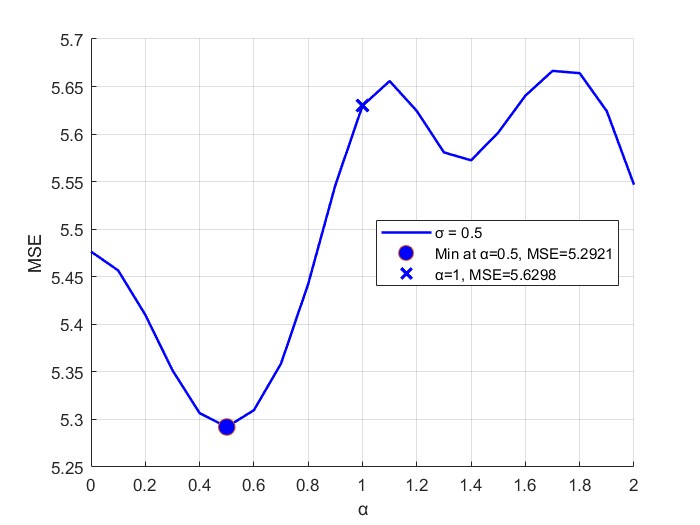}\label{fig:sub10}}\hfill
\subfloat[]{\includegraphics[width=0.32\textwidth]{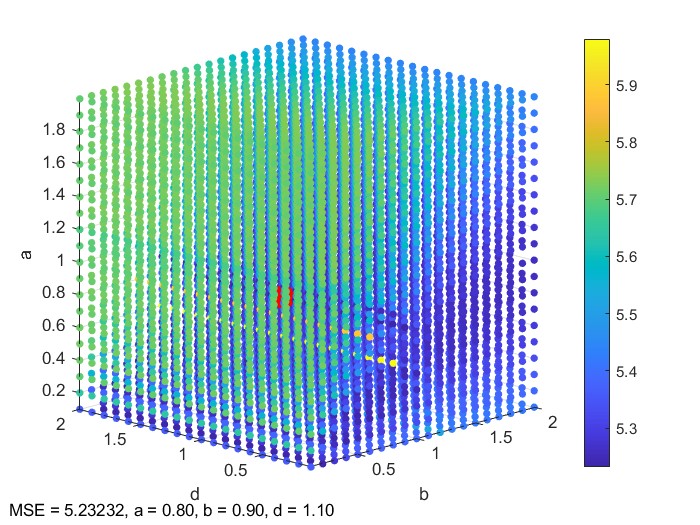}\label{fig:sub11}}\hfill
\subfloat[]{\includegraphics[width=0.32\textwidth]{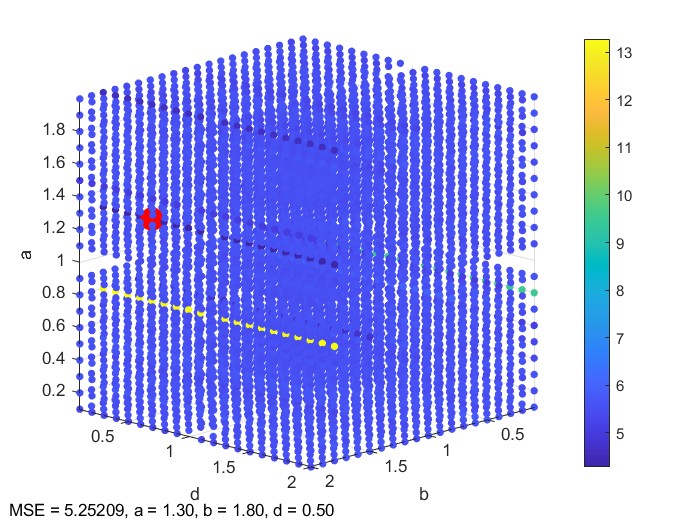}\label{fig:sub12}}
\caption{Denoising results (MSE) on a Swiss roll graph ($s = 0.5$) : comparison of GFRFT and GLCT under different GSOs. (a)-(c) use the weighted adjacency matrix; (d)-(f) use the Laplacian matrix; b: wAdj-CDDHFs-GLCT; c: wAdj-CM-CC-CM-GLCT; e: Lap-CDDHFs-GLCT; f: Lap-CM-CC-CM-GLCT.}
\label{fig:Swiss roll-MSE}
\end{figure*}

\begin{figure*}[!t]
\centering
\subfloat[]{\includegraphics[width=0.32\textwidth]{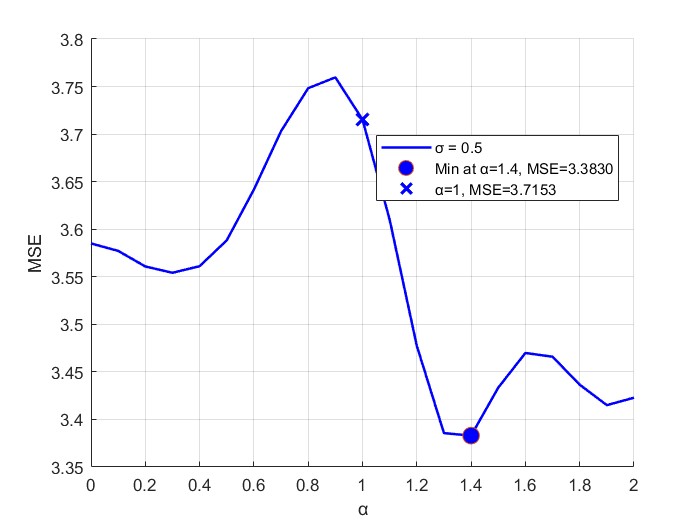}\label{fig:sub13}}\hfill
\subfloat[]{\includegraphics[width=0.32\textwidth]{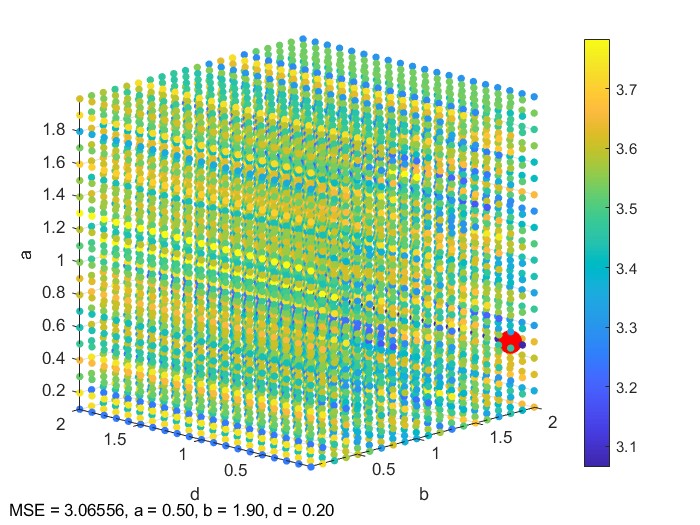}\label{fig:sub14}}\hfill
\subfloat[]{\includegraphics[width=0.32\textwidth]{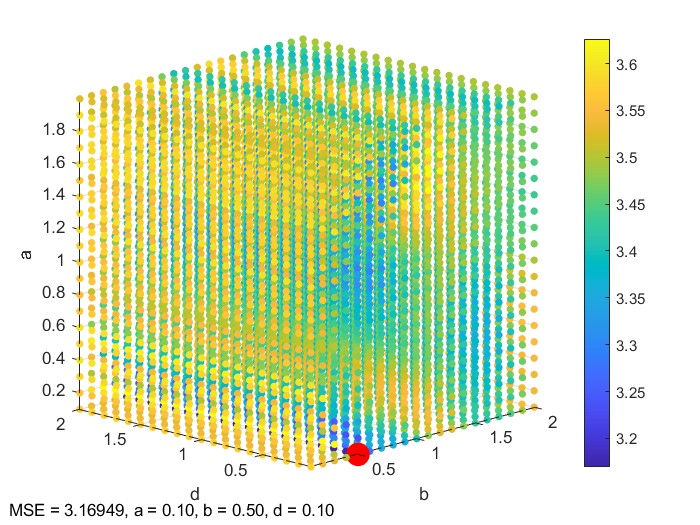}\label{fig:sub15}}\\[0.5em]
\subfloat[]{\includegraphics[width=0.32\textwidth]{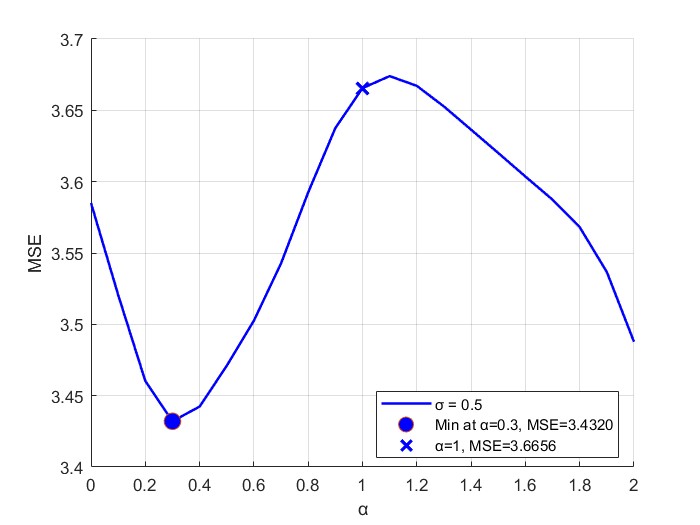}\label{fig:sub16}}\hfill
\subfloat[]{\includegraphics[width=0.32\textwidth]{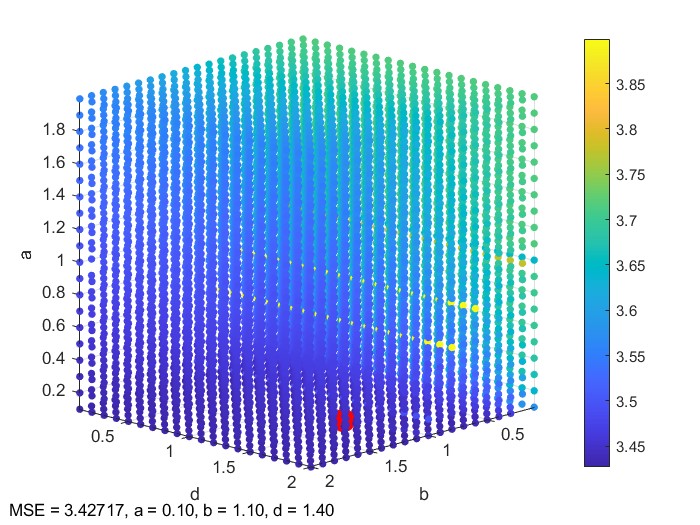}\label{fig:sub17}}\hfill
\subfloat[]{\includegraphics[width=0.32\textwidth]{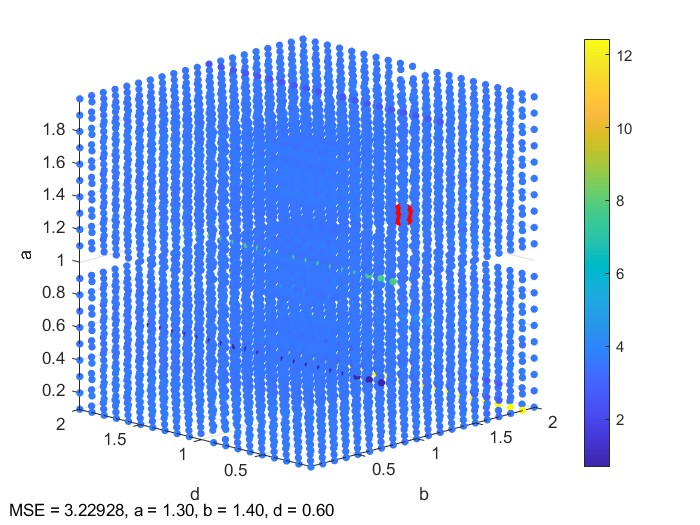}\label{fig:sub18}}
\caption{Denoising results (MSE) on a sensor graph ($s = 0.5$) : comparison of GFRFT and GLCT under different GSOs. (a)-(c) use the weighted adjacency matrix; (d)-(f) use the Laplacian matrix; b: wAdj-CDDHFs-GLCT; c: wAdj-CM-CC-CM-GLCT; e: Lap-CDDHFs-GLCT; f: Lap-CM-CC-CM-GLCT.}
\label{fig:Sensor-MSE}
\end{figure*}

The results in Tables \ref{tab:denoising_results_5nn}--\ref{tab:denoising_results_sensor}, demonstrate that the GLCT methods consistently achieve superior denoising performance compared to both conventional GFT and GFRFT methods under the same eigenbasis framework. The visual comparisons at the moderate noise level ($s=0.5$), as shown in Figs. \ref{fig:5-nn-MSE}--\ref{fig:Sensor-MSE}. The traditional GFT approach exhibits noticeably inferior results, highlighting the inherent limitations of fixed basis transforms in handling complex graph signals. Although the GFRFT method improves flexibility through the introduction of fractional orders, its performance remains significantly behind that of the GLCT method. As the noise intensity increases from $s = 0.5$ to $1.5$, the MSE values of all methods show the expected upward trend. The GLCT method maintains a consistent relative performance advantage throughout. These results indicate that GLCT delivers the best denoising performance across various graph structures and noise levels.
\subsection{Real-World Data}
\begin{figure*}[!t]
    \centering
    \begin{minipage}{0.3\textwidth}
        \centering
        \includegraphics[width=\linewidth]{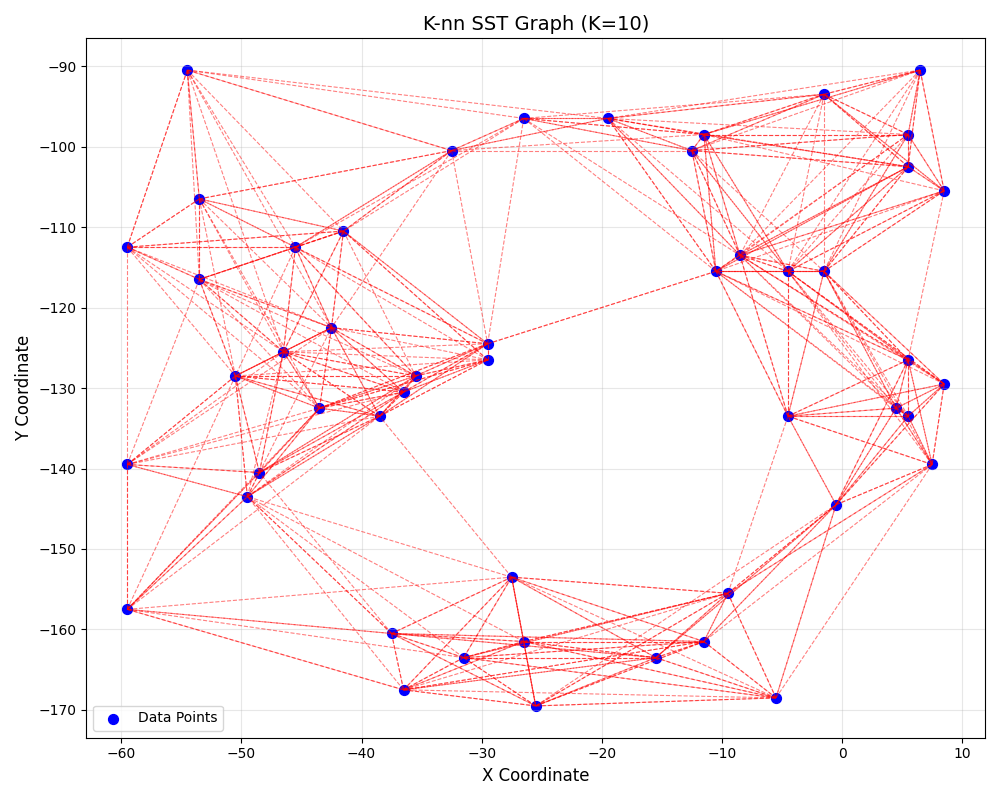}
        \caption{10-nn SST graph with 50 nodes.}\label{fig:10-nnSST}
    \end{minipage}\hfill
    \begin{minipage}{0.3\textwidth}
        \centering
        \includegraphics[width=\linewidth]{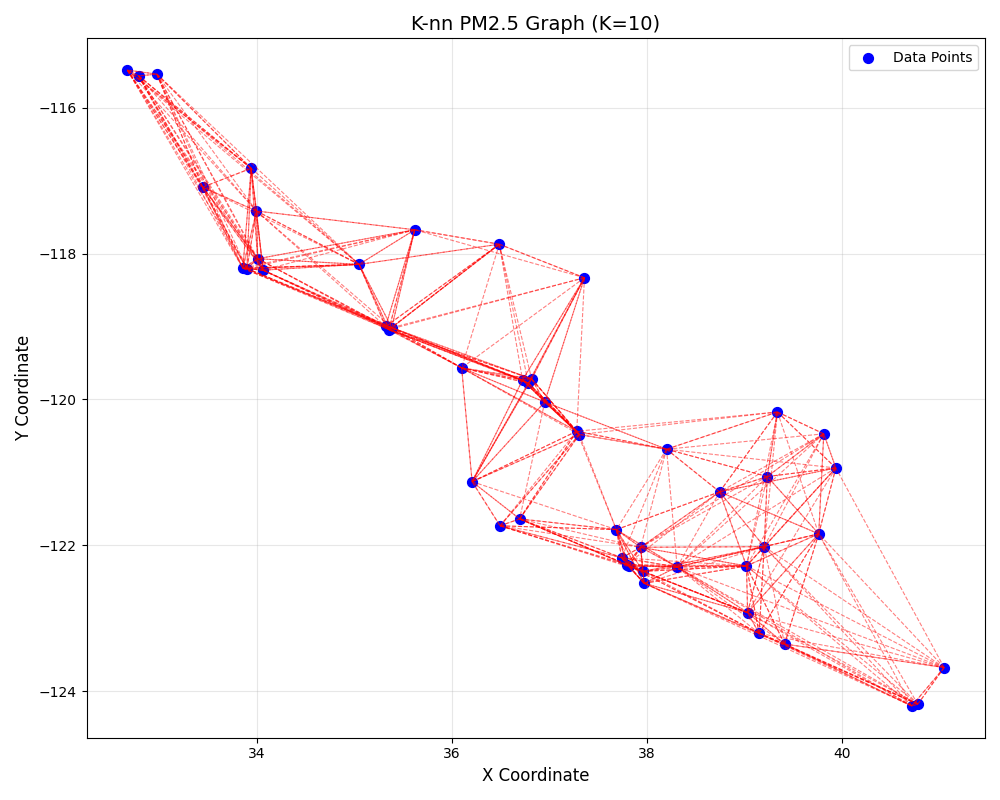}
        \caption{10-nn PM2.5 graph with 50 nodes.}\label{fig:10-nnPM2.5}
    \end{minipage}\hfill
    \begin{minipage}{0.3\textwidth}
        \centering
        \includegraphics[width=\linewidth]{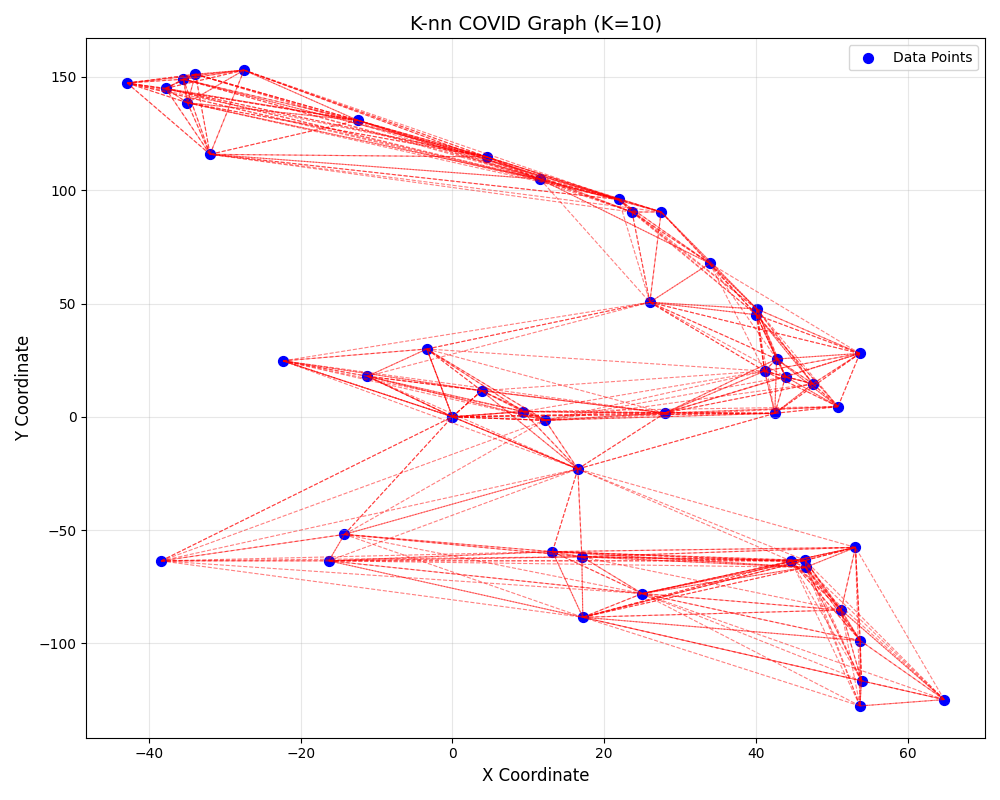}
        \caption{10-nn COVID graph with 50 nodes.}\label{fig:10-nnCOVID}
    \end{minipage}
\end{figure*}

We constructed graphs from three datasets: Sea Surface Temperature (SST), Particulate Matter 2.5 (PM2.5), and COVID-19 (COVID)\cite{33}. The underlying graph is not predetermined, we use $k$-nn graphs for three $k$ values: $k \in \{2,6,10\}$, each consisting of $N =50$ nodes, as shown in Figs. \ref{fig:10-nnSST}--\ref{fig:10-nnCOVID}. For each dataset, we evaluate observations at three time points $t_1,t_2,t_3$, treating them as independent snapshots on the same graph. To assess robustness, the clean signals are perturbed with additive zero-mean Gaussian noise at multiple intensities, with standard deviation $s \in \{0.5,0.6,0.7\}$. 

Given the large number of nodes in real-world graphs, we employ an Adam-based joint optimization strategy to simultaneously learn the parameters of the GLCT and the diagonal filter, which mitigates the combinatorial curse of large-scale grid search. The experimental settings are as follows:
\begin{itemize}
    \item Adam-based joint optimization: learning rate: $\varepsilon = 0.005$; iterations: 5000.
    \item Initialization: all trainable parameters are initialized i.i.d. as
    $\theta \sim \mathcal{N}(0,1)$; random seed fixed.
\end{itemize}

The results in Tables \ref{tab:denoising_results_SST}--\ref{tab:denoising_results_COVID} show that denoising MSE on three real-world datasets under a complete grid of graph densities $k \in \{2,6,10\}$ and noise levels $s\in\{0.5,0.6,0.7\}$. Across datasets and settings, GLCT methods achieve the best or tied-best performance. For example, in the SST dataset, \Cref{tab:denoising_results_SST} shows that the wAdj-CDDHFs-GLCT method achieved the lowest MSE value of 1.44195 under the $k=2,s=0.5$ setting, significantly outperforming traditional GFRFT methods, representing approximately 25\%. In terms of noise robustness, as $s$ increases from 0.5 to 0.7, all approaches experience higher MSE, yet GLCT variants degrade more gracefully. The performance improvement can be largely attributed to the additional degrees of freedom offered by the GLCT, which enables a signal representation that more effectively concentrates key information. This facilitates noise removal while better preserving semantically important features. By simultaneously adjusting both the transform parameters and filter coefficients via Adam-based joint optimization, the method achieves more stable performance across diverse datasets, which effectively reduces the mismatch commonly caused by the conventional ``transform-then-filter" pipelines.

A closer examination across graph densities clarifies the conditions for performance gains. Transitioning from sparse graphs to moderately connected ones generally enhances reconstruction quality, as sufficient neighborhood information stabilizes the spectral domain for more reliable filtering. However, as the number of edges further increases, the marginal benefits begin to diminish. Especially for signals dominated by fine-scale structure, overly dense connectivity tends to induce over-smoothing, thereby weakening the preservation of high-frequency details. Accordingly, the choice of graph density should balance the need for sufficient structural information against the ability to preserve details.

An in-depth investigation is conducted to elucidate the intrinsic mechanisms underlying the exceptional performance of the GLCT methods. Their core advantage stems from the data-driven nature of the approach, which employs an adaptive learning mechanism to identify the optimal joint vertex-spectral representation for specific graph signal datasets, thereby enabling more precise separation of signal from noise. The joint optimization strategy adopted in this study also demonstrates value. By simultaneously learning all parameters of the transform operator and the filter within a unified framework, this approach achieves end-to-end co-adaptation and effectively mitigates the high computational complexity inherent in traditional sequential parameter-tuning methods. This not only substantially reduces the computational burden and eliminates the need for multiple iterations and manual intervention, but also significantly enhances the stability and convergence efficiency of the overall optimization process.


\section{Conclusion}
In this paper, we presented a learnable GLCT-GWF that jointly optimizes the transform parameters and a diagonal spectral filter under a unified MSE objective. Our approach fuses domain selection and filtering into a single differentiable procedure, eliminating the combinatorial burden of grid search and reducing computational cost in practice. Across synthetic data and three real-world datasets (SST, PM2.5, and COVID), GLCT-based denoising consistently outperforms GFRFT-based counterparts. Essentially, the advantage stems from the synergistic effect between the richer parameterization of the GLCT and joint learning: the former adaptively optimizes the signal representation in the transform domain, achieving clearer separation between signal and noise; the latter avoids the suboptimal results and biases that may arise from independent step-by-step optimization by enabling end-to-end joint optimization of the transform and filtering process. Experiments demonstrated that GLCT-GWF consistently improves signal-noise separation across diverse graph topologies and noise conditions, outperforming GFRFT-based denoisers. These gains confirmed the robustness of joint transform learning transform domain selection, indicating a scalable and effective solution for large-scale GSP.
\begingroup
\setlength{\tabcolsep}{6pt}        % 列间距：稍微更大
\renewcommand{\arraystretch}{2.05}  % 行距：稍微更大
\setlength{\aboverulesep}{3pt}      % 规则线上下留白：略增
\setlength{\belowrulesep}{3pt}

\begin{table*}[!t]
  \centering
  \footnotesize
  \caption{Denoising Results (MSE) on the SST Dataset}
  \label{tab:sst_mse}
  % 每列统一的 sigma 文本（小号）
  \newcommand{\sig}{\scriptsize$s{=}0.5/0.6/0.7$}
  \resizebox{\textwidth}{!}{
  \begin{tabular}{c *{9}{c}}
    \toprule
    \multirow{3}{*}{\textbf{Method}}
      & \multicolumn{3}{c}{\textbf{2-nn}}
      & \multicolumn{3}{c}{\textbf{6-nn}}
      & \multicolumn{3}{c}{\textbf{10-nn}} \\
    \cmidrule(lr){2-4}\cmidrule(lr){5-7}\cmidrule(lr){8-10}
      %—— 每个列头两行：t=... + sigma（把原来的 \\[-1pt] 改为普通 \\）
      & \makecell{$t_1{=}200$\\ \sig}
      & \makecell{$t_2{=}500$\\ \sig}
      & \makecell{$t_3{=}800$\\ \sig}
      & \makecell{$t_1{=}200$\\ \sig}
      & \makecell{$t_2{=}500$\\ \sig}
      & \makecell{$t_3{=}800$\\ \sig}
      & \makecell{$t_1{=}200$\\ \sig}
      & \makecell{$t_2{=}500$\\ \sig}
      & \makecell{$t_3{=}800$\\ \sig} \\
    \midrule

    % ===== 把每格第二行的 \\[-2pt] 改成 \\[1pt]（略微放大） =====
    GFRFT$_\mathbf{A}$
      & \makecell{2.62332/3.38724/4.20664\\[1pt]}
      & \makecell{2.26568/3.03248/3.85832\\[1pt]}
      & \makecell{2.63676/3.41272/4.24255\\[1pt]}
      & \makecell{0.99828/1.31640/1.67586\\[1pt]}
      & \makecell{1.05027/1.39152/1.77623\\[1pt]}
      & \makecell{1.00383/1.32278/1.68295\\[1pt]}
      & \makecell{3.22283/3.98299/4.76354\\[1pt]}
      & \makecell{3.28789/4.39580/5.61469\\[1pt]}
      & \makecell{3.28489/4.06487/4.85269\\[1pt]}
    \\
    GFRFT$_\mathbf{L}$
      & \makecell{3.33222/4.31053/5.31161\\[1pt]}
      & \makecell{2.44334/3.39352/4.44257\\[1pt]}
      & \makecell{2.29709/3.16895/4.13145\\[1pt]}
      & \makecell{0.47760/0.62449/0.79428\\[1pt]}
      & \makecell{0.42973/0.59696/0.78783\\[1pt]}
      & \makecell{0.41026/0.56785/0.74200\\[1pt]}
      & \makecell{0.44281/0.62064/0.82437\\[1pt]}
      & \makecell{0.44355/0.62435/0.83214\\[1pt]}
      & 
      \makecell{0.43688/0.61448/0.81775\\[1pt]}
    \\
    wAdj-CDDHFs-GLCT
      & \makecell{\textbf{1.44195/2.01209/2.64687}\\[1pt]} 
      & \makecell{\textbf{1.43019/1.97336/2.57752}\\[1pt]}
      & \makecell{\textbf{1.45264/2.01930/2.64559}\\[1pt]}
      & \makecell{0.70033/0.98078/1.30666\\[1pt]}
      & \makecell{0.71799/1.02123/1.37290\\[1pt]}
      & \makecell{0.67799/0.96278/1.29292\\[1pt]}
      & \makecell{3.14113/3.95131/4.76350\\[1pt]}
      & \makecell{2.35597/3.13809/4.00430\\[1pt]}
      & \makecell{2.82555/3.68102/4.61852\\[1pt]}
    \\
      Lap-CDDHFs-GLCT
      & \makecell{2.33613/3.11043/3.84672\\[1pt]}
      & \makecell{1.94344/2.60104/3.28401\\[1pt]}
      & \makecell{2.09507/2.77444/3.47567\\[1pt]}
      & \makecell{\textbf{0.45256/0.61349/0.79413}\\[1pt]}
      & \makecell{\textbf{0.42404/0.59137/0.78319}\\[1pt]}
      & \makecell{\textbf{0.41025/0.56480/0.74197}\\[1pt]}
      & \makecell{\textbf{0.44276/0.62064/0.82437}\\[1pt]}
      & \makecell{\textbf{0.44355/0.62435/0.83214}\\[1pt]}
      & \makecell{\textbf{0.43683/0.61423/0.81764}\\[1pt]}
    \\
      wAdj-CM-CC-CM-GLCT
      & \makecell{1.98174/2.72336/3.53919\\[1pt]}
      & \makecell{1.89208/2.60585/3.39559\\[1pt]}
      & \makecell{1.99334/2.74179/3.56551\\[1pt]}
      & \makecell{2.88051/3.71440/4.63524\\[1pt]}
      & \makecell{2.88049/3.79813/4.68590\\[1pt]}
      & \makecell{2.99534/3.70514/4.61575\\[1pt]}
      & \makecell{2.94778/3.69479/4.50985\\[1pt]}
      & \makecell{3.11771/3.93379/4.82982\\[1pt]}
      & \makecell{2.77807/3.55315/4.39887\\[1pt]}
    \\
      Lap-CM-CC-CM-GLCT
      & \makecell{2.04535/2.87299/3.80316\\[1pt]}
      & \makecell{2.08320/2.92538/3.73083\\[1pt]}
      & \makecell{2.07221/2.89442/3.6484\\[1pt]}
      & \makecell{0.71531/1.00201/1.33594\\[1pt]}
      & \makecell{0.73022/1.03853/1.39701\\[1pt]}
      & \makecell{0.70332/0.99157/1.32656\\[1pt]}
      & \makecell{0.49741/0.71099/0.95965\\[1pt]}
      & \makecell{0.51651/0.73863/0.99742\\[1pt]}
      & \makecell{0.50165/0.71716/0.96812\\[1pt]}
    \\
    % 你后面如果还有重复的 GFT_A 行，同样把 \\[-2pt] 全部改为 \\[1pt]
    \bottomrule
  \end{tabular}}
  \label{tab:denoising_results_SST}
\end{table*}

\endgroup

\begingroup
\setlength{\tabcolsep}{6pt}        % 列间距：稍微更大
\renewcommand{\arraystretch}{2.05}  % 行距：稍微更大
\setlength{\aboverulesep}{3pt}      % 规则线上下留白：略增
\setlength{\belowrulesep}{3pt}

\begin{table*}[!t]
  \centering
  \footnotesize
  \caption{Denoising Results (MSE) on the PM2.5 Dataset}
  \label{tab:pm_mse}
  % 每列统一的 sigma 文本（小号）
  \newcommand{\sig}{\scriptsize$s{=}0.5/0.6/0.7$}
  \resizebox{\textwidth}{!}{
  \begin{tabular}{c *{9}{c}}
    \toprule
    \multirow{3}{*}{\textbf{Method}}
      & \multicolumn{3}{c}{\textbf{2-nn}}
      & \multicolumn{3}{c}{\textbf{6-nn}}
      & \multicolumn{3}{c}{\textbf{10-nn}} \\
    \cmidrule(lr){2-4}\cmidrule(lr){5-7}\cmidrule(lr){8-10}
      %—— 每个列头两行：t=... + sigma（把原来的 \\[-1pt] 改为普通 \\）
      & \makecell{$t_1{=}50$\\ \sig}
      & \makecell{$t_2{=}125$\\ \sig}
      & \makecell{$t_3{=}200$\\ \sig}
      & \makecell{$t_1{=}50$\\ \sig}
      & \makecell{$t_2{=}125$\\ \sig}
      & \makecell{$t_3{=}200$\\ \sig}
      & \makecell{$t_1{=}50$\\ \sig}
      & \makecell{$t_2{=}125$\\ \sig}
      & \makecell{$t_3{=}200$\\ \sig} \\
    \midrule

    % ===== 把每格第二行的 \\[-2pt] 改成 \\[1pt]（略微放大） =====
    GFRFT$_\mathbf{A}$
      & \makecell{1.27485/1.53299/1.79042\\[1pt]}
      & \makecell{3.44867/4.27995/5.14405\\[1pt]}
      & \makecell{2.87846/3.62801/4.40876\\[1pt]}
      & \makecell{0.72415/0.95847/1.19603\\[1pt]}
      & \makecell{3.36142/4.27005/5.13739\\[1pt]}
      & \makecell{3.29393/4.29362/5.27401\\[1pt]}
      & \makecell{1.52852/1.83730/2.15620\\[1pt]}
      & \makecell{4.48982/5.44981/6.38206\\[1pt]}
      & \makecell{2.32751/2.76995/3.00690\\[1pt]}
    \\
    GFRFT$_\mathbf{L}$
      & \makecell{1.24380/1.52780/1.81383\\[1pt]}
      & \makecell{4.60015/5.29349/6.00934\\[1pt]}
      & \makecell{3.29303/4.24019/5.21014\\[1pt]}
      & \makecell{2.22684/2.56617/2.90555\\[1pt]}
      & \makecell{2.98594/3.69954/4.47705\\[1pt]}
      & \makecell{3.28737/3.81018/4.34654\\[1pt]}
      & \makecell{0.93559/1.89633/2.17899\\[1pt]}
      & \makecell{2.24587/2.91711/3.62668\\[1pt]}
      & \makecell{1.36483/1.86987/2.42000\\[1pt]}
    \\
    wAdj-CDDHFs-GLCT
      & \makecell{0.66268/0.86365/1.07927\\[1pt]} 
      & \makecell{\textbf{2.15787/2.73239/3.30814}\\[1pt]}
      & \makecell{4.48757/5.90064/7.33773\\[1pt]}
      & \makecell{0.47477/0.63283/0.80264\\[1pt]}
      & \makecell{\textbf{1.87638/2.56117/3.32083}\\[1pt]}
      & \makecell{2.77561/3.29275/4.19509\\[1pt]}
      & \makecell{0.61789/0.81186/1.01759\\[1pt]}
      & \makecell{4.20963/5.15709/6.04754\\[1pt]}
      & \makecell{2.12743/2.55182/3.00691\\[1pt]}
    \\
      Lap-CDDHFs-GLCT
      & \makecell{\textbf{0.47004/0.62098/0.78921}\\[1pt]}
      & \makecell{3.74040/4.63850/5.53651\\[1pt]}
      & \makecell{2.62230/3.28516/4.00987\\[1pt]}
      & \makecell{1.42938/1.82389/2.23712\\[1pt]}
      & \makecell{2.75983/3.20182/3.64193\\[1pt]}
      & \makecell{2.55936/3.09910/3.70015\\[1pt]}
      & \makecell{0.58933/1.35561/1.78109\\[1pt]}
      & \makecell{\textbf{1.40680/1.88933/2.43525}\\[1pt]}
      & \makecell{1.18297/1.63149/2.12752\\[1pt]}
    \\
      wAdj-CM-CC-CM-GLCT
      & \makecell{0.57365/0.75954/0.95876\\[1pt]}
      & \makecell{2.39282/3.05124/3.73926\\[1pt]}
      & \makecell{2.84944/3.59859/4.37966\\[1pt]}
      & \makecell{\textbf{0.47476/0.63283/0.80264}\\[1pt]}
      & \makecell{4.11521/5.02166/5.95282\\[1pt]}
      & \makecell{2.56417/3.00856/3.47965\\[1pt]}
      & \makecell{\textbf{0.56935/0.77214/0.93609}\\[1pt]}
      & \makecell{3.56511/4.97070/5.30724\\[1pt]}
      & \makecell{2.59766/3.21214/3.95942\\[1pt]}
    \\
      Lap-CM-CC-CM-GLCT
      & \makecell{1.14327/1.47171/1.80973\\[1pt]}
      & \makecell{2.40138/3.12124/3.87158\\[1pt]}
      & \makecell{\textbf{2.22716/2.88266/3.54164}\\[1pt]}
      & \makecell{1.33013/1.71737/2.11042\\[1pt]}
      & \makecell{2.26044/2.94418/3.64093\\[1pt]}
      & \makecell{\textbf{1.95329/2.52835/3.10474}\\[1pt]}
      & \makecell{0.71372/0.97688/1.25948\\[1pt]}
      & \makecell{1.59843/2.13246/2.71513\\[1pt]}
      & \makecell{\textbf{1.00874/1.30875/1.62981}\\[1pt]}
    \\
    % 你后面如果还有重复的 GFT_A 行，同样把 \\[-2pt] 全部改为 \\[1pt]
    \bottomrule
  \end{tabular}}
  \label{tab:denoising_results_PM}
\end{table*}

\endgroup

\begingroup
\setlength{\tabcolsep}{6pt}        % 列间距：稍微更大
\renewcommand{\arraystretch}{2.05}  % 行距：稍微更大
\setlength{\aboverulesep}{3pt}      % 规则线上下留白：略增
\setlength{\belowrulesep}{3pt}

\begin{table*}[!t]
  \centering
  \footnotesize
  \caption{Denoising Results (MSE) on the COVID Dataset}
  \label{tab:covid_mse}
  % 每列统一的 sigma 文本（小号）
  \newcommand{\sig}{\scriptsize$s{=}0.5/0.6/0.7$}
  \resizebox{\textwidth}{!}{
  \begin{tabular}{c *{9}{c}}
    \toprule
    \multirow{3}{*}{\textbf{Method}}
      & \multicolumn{3}{c}{\textbf{2-nn}}
      & \multicolumn{3}{c}{\textbf{6-nn}}
      & \multicolumn{3}{c}{\textbf{10-nn}} \\
    \cmidrule(lr){2-4}\cmidrule(lr){5-7}\cmidrule(lr){8-10}
      %—— 每个列头两行：t=... + sigma（把原来的 \\[-1pt] 改为普通 \\）
      & \makecell{$t_1{=}50$\\ \sig}
      & \makecell{$t_2{=}80$\\ \sig}
      & \makecell{$t_3{=}110$\\ \sig}
      & \makecell{$t_1{=}50$\\ \sig}
      & \makecell{$t_2{=}80$\\ \sig}
      & \makecell{$t_3{=}110$\\ \sig}
      & \makecell{$t_1{=}50$\\ \sig}
      & \makecell{$t_2{=}80$\\ \sig}
      & \makecell{$t_3{=}110$\\ \sig} \\
    \midrule

    % ===== 把每格第二行的 \\[-2pt] 改成 \\[1pt]（略微放大） =====
    GFRFT$_\mathbf{A}$
      & \makecell{2.18520/2.84692/3.54282\\[1pt]}
      & \makecell{3.99323/5.37056/6.89006\\[1pt]}
      & \makecell{1.78537/2.26996/2.82572\\[1pt]}
      & \makecell{5.86263/6.59129/7.40987\\[1pt]}
      & \makecell{4.05156/5.39565/6.82562\\[1pt]}
      & \makecell{3.91232/4.62640/5.38477\\[1pt]}
      & \makecell{1.83720/2.56281/3.37117\\[1pt]}
      & \makecell{2.01447/2.33798/3.05043\\[1pt]}
      & \makecell{1.88387/2.63449/3.47051\\[1pt]}
    \\
    GFRFT$_\mathbf{L}$
      & \makecell{1.16289/1.61632/2.13446\\[1pt]}
      & \makecell{3.16254/4.39540/5.81172\\[1pt]}
      & \makecell{1.83519/2.41501/3.02088\\[1pt]}
      & \makecell{1.37538/1.89021/2.38548\\[1pt]}
      & \makecell{2.48550/2.87302/3.31348\\[1pt]}
      & \makecell{3.50878/4.71889/5.96887\\[1pt]}
      & \makecell{1.23290/1.69771/2.21065\\[1pt]}
      & \makecell{4.98030/6.40701/8.03110\\[1pt]}
      & 
      \makecell{3.08944/3.898802/4.70447\\[1pt]}
    \\
    wAdj-CDDHFs-GLCT
      & \makecell{1.38602/1.98639/2.68935\\[1pt]} 
      & \makecell{1.93471/2.30486/2.81012\\[1pt]}
      & \makecell{\textbf{1.34312/1.75051/2.19421}\\[1pt]}
      & \makecell{3.93558/5.35085/6.89180\\[1pt]}
      & \makecell{3.28279/3.91168/4.56482\\[1pt]}
      & \makecell{1.54002/2.15675/2.85440\\[1pt]}
      & \makecell{1.40349/1.71742/2.31381\\[1pt]}
      & \makecell{\textbf{0.80017/1.13266/1.51601}\\[1pt]}
      & \makecell{\textbf{1.10936/1.57256/2.10251}\\[1pt]}
    \\
      Lap-CDDHFs-GLCT
      & \makecell{\textbf{0.78027/1.08142/1.42493}\\[1pt]}
      & \makecell{1.36238/1.82520/2.32761\\[1pt]}
      & \makecell{1.46576/2.06482/2.74671\\[1pt]}
      & \makecell{\textbf{0.76959/1.07510/1.42508}\\[1pt]}
      & \makecell{\textbf{1.18489/1.61181/2.10564}\\[1pt]}
      & \makecell{2.03941/2.46971/2.93865\\[1pt]}
      & \makecell{\textbf{0.89279/1.26077/1.65955}\\[1pt]}
      & \makecell{2.00565/2.60243/3.22813\\[1pt]}
      & \makecell{1.55736/2.42499/3.15737\\[1pt]}
    \\
      wAdj-CM-CC-CM-GLCT
      & \makecell{1.33589/1.90279/2.52269\\[1pt]}
      & \makecell{3.03960/4.00468/5.07900\\[1pt]}
      & \makecell{1.36691/1.89941/2.49521\\[1pt]}
      & \makecell{1.13594/1.59428/2.08279\\[1pt]}
      & \makecell{4.83931/4.97147/6.20468\\[1pt]}
      & \makecell{\textbf{1.45915/2.01452/2.57427}\\[1pt]}
      & \makecell{0.89791/1.26547/1.70285\\[1pt]}
      & \makecell{2.97678/3.88895/4.38408\\[1pt]}
      & \makecell{1.35960/1.93219/2.58995\\[1pt]}
    \\
      Lap-CM-CC-CM-GLCT
      & \makecell{0.93234/1.28030/1.66140\\[1pt]}
      & \makecell{\textbf{1.23983/1.72032/2.25439}\\[1pt]}
      & \makecell{1.54181/2.08195/2.65633\\[1pt]}
      & \makecell{1.19500/1.68667/2.24772\\[1pt]}
      & \makecell{1.20929/1.67193/2.19448\\[1pt]}
      & \makecell{2.04535/2.87299/3.80316\\[1pt]}
      & \makecell{1.13953/1.56592/2.04056\\[1pt]}
      & \makecell{2.00600/5.60295/3.22903\\[1pt]}
      & \makecell{2.08630/2.47398/2.90491\\[1pt]}
    \\
    % 你后面如果还有重复的 GFT_A 行，同样把 \\[-2pt] 全部改为 \\[1pt]
    \bottomrule
  \end{tabular}}
  \label{tab:denoising_results_COVID}
\end{table*}
\endgroup
  
\appendices
\section{Proofs of Properties 3--5}
\subsubsection*{Proof of Property 3 (Additivity)}
Let $\mathbf{M}_j = (a_j,b_j;c_j,d_j)$ and
$(\xi_{1,j},\xi_{2,j},\xi_{3,j})=(\frac{d_j-1}{b_j},-b_j,\frac{a_j-1}{b_j})$ for $j=1,2$.
Expanding the definition gives the six-factor chain
\begin{equation}
\begin{aligned}
\mathcal{F}^{\mathbf{M}_2}_{\mathbf{L}}\mathcal{F}^{\mathbf{M}_1}_{\mathbf{L}}
&= \mathcal{CM}^{\xi_{1,2}}_{\mathbf{L}}\; \mathbf{U}_{\mathbf{L}} \mathbf{J}^{\xi_{2,2}}_{\mathbf{L}} \mathbf{U}_{\mathbf{L}}^{-1}\; \mathcal{CM}^{\xi_{3,2}}_{\mathbf{L}} \\
&\quad \mathcal{CM}^{\xi_{1,1}}_{\mathbf{L}}\; \mathbf{U}_{\mathbf{L}} \mathbf{J}^{\xi_{2,1}}_{\mathbf{L}} \mathbf{U}_{\mathbf{L}}^{-1}\; \mathcal{CM}^{\xi_{3,1}}_{\mathbf{L}},
\end{aligned}
\end{equation}
where $\mathcal{CM}^{\xi_{2,2}}_{\mathbf{L}}$ and $\mathcal{CM}^{\xi_{2,1}}_{\mathbf{L}}$ are diagonal matrices, here we use $\mathbf{J}^{\xi_{2,2}}_{\mathbf{L}}$ and $\mathbf{J}^{\xi_{2,1}}_{\mathbf{L}}$ for representation.

\indent\textbf{Step 1: Move the Two GCMs into the Spectral Sandwich.}

Insert the identity $\mathbf{U}_{\mathbf{L}}\mathbf{U}_{\mathbf{L}}^{-1}=\mathbf{I}$ between the two GCC blocks and regroup:
{\setlength{\abovedisplayskip}{6pt}% 公式上方与文本的间距
 \setlength{\belowdisplayskip}{6pt}% 公式下方与文本的间距
 \setlength{\abovedisplayshortskip}{2pt}% 短公式（无编号等情况）上方间距
 \setlength{\belowdisplayshortskip}{2pt}% 短公式下方间距
\begin{equation}
\begin{aligned}
\mathcal{F}^{\mathbf{M}_2}_{\mathbf{L}}\mathcal{F}^{\mathbf{M}_1}_{\mathbf{L}}
= \mathcal{CM}^{\xi_{1,2}}_{\mathbf{L}}\,\mathbf{U}_{\mathbf{L}}\;
   \underbrace{\Big(\cdots\Big)}_{\textbf{middle block}}  \mathbf{U}_{\mathbf{L}}^{-1}\,\mathcal{CM}^{\xi_{3,1}}_{\mathbf{L}},
\label{equ:附录A公式1}
\end{aligned}
\end{equation}}
\noindent where the ``middle block'' is specifically equal to $\mathbf{J}^{\xi_{2,2}}_{\mathbf{L}}\,
   \mathbf{U}_{\mathbf{L}}^{-1}\mathcal{CM}^{\xi_{3,2}}_{\mathbf{L}}\mathbf{U}_{\mathbf{L}}\,
   \mathbf{U}_{\mathbf{L}}^{-1}\mathcal{CM}^{\xi_{1,1}}_{\mathbf{L}}\mathbf{U}_{\mathbf{L}}\,
   \mathbf{J}^{\xi_{2,1}}_{\mathbf{L}}$.

\indent\textbf{Step 2: Absorb the ``Middle Block'' into a New Spectral Phase.}

Viewing the bracket  as a consolidated spectral operation, there exist a unique
$\xi_2'\in\mathbb{R}$ such that
\begin{equation}
\Big(\mathbf{J}^{\xi_{2,2}}_{\mathbf{L}}\,
   \mathbf{U}_{\mathbf{L}}^{-1}\mathcal{CM}^{\xi_{3,2}}_{\mathbf{L}}\mathbf{U}_{\mathbf{L}}\,
   \mathbf{U}_{\mathbf{L}}^{-1}\mathcal{CM}^{\xi_{1,1}}_{\mathbf{L}}\mathbf{U}_{\mathbf{L}}\,
   \mathbf{J}^{\xi_{2,1}}_{\mathbf{L}}\Big) \;=\; \mathcal{CM}^{\xi_2'}_{\mathbf{L}}.
   \label{equ:附录A公式2}
\end{equation}

Intuitively, the two moved factors $\mathbf{U}_{\mathbf{L}}^{-1}\mathcal{CM}_{\mathbf{L}}\mathbf{U}_{\mathbf{L}}$ are absorbed together with the
boundary spectral phases $\mathbf{J}^{\xi_{2,2}}_{\mathbf{L}}$ and $\mathbf{J}^{\xi_{2,1}}_{\mathbf{L}}$ into a single GCM.

\indent\textbf{Step 3: Return to CM-CC-CM Form and Read Parameters.}

Substituting \eqref{equ:附录A公式2} back into \eqref{equ:附录A公式1} and comparing with the definition yields
\begin{equation}
\mathcal{F}^{\mathbf{M}_2}_{\mathbf{L}}\mathcal{F}^{\mathbf{M}_1}_{\mathbf{L}}
\;=\;\mathcal{CM}^{\xi_{1,2}}_{\mathbf{L}}\,\mathbf{U}_{\mathbf{L}} \mathcal{CM}^{\xi_2'}_{\mathbf{L}} \mathbf{U}_{\mathbf{L}}^{-1}\,\mathcal{CM}^{\xi_{3,1}}_{\mathbf{L}}.
\end{equation}
There exists a unique
$\mathbf{M}'=\begin{bmatrix}a'&b'\\ c'&d'\end{bmatrix}\in \mathrm{SL}(2,\mathbb{R})$ (with $b'\neq0$) such that
\begin{equation}
    (\xi_1',\xi_2',\xi_3')=\Big(\frac{d'}{b'},\ -b',\ \frac{a'}{b'}\Big).
\end{equation}
By the one-to-one correspondence between CM-CC-CM parameters and
$2\times2$ matrices in $\mathrm{SL}(2,\mathbb{R})$, the matrix must be the product $\mathbf{M'=M_2M_1}$; 
\begin{equation}
\mathcal{F}^{\mathbf{M}_2}_{\mathbf{L}}\mathcal{F}^{\mathbf{M}_1}_{\mathbf{L}}\;=\;\mathcal{F}^{\mathbf{M}_2\mathbf{M}_1}_{\mathbf{L}}.
\end{equation}

\subsubsection*{Proof of Property 4 (Invertibility)}
The proof of invertibility follows directly from the additivity property established previously. Setting $\mathbf{M_2}=\mathbf{M_1}^{-1}$ immediately yields 
\begin{equation} \mathcal{F}^{\mathbf{M}_2}_{\mathbf{L}}\mathcal{F}^{\mathbf{M}_1}_{\mathbf{L}}=\mathcal{F}^{\mathbf{M_1}^{-1}}_{\mathbf{L}}\mathcal{F}^{\mathbf{M}_1}_{\mathbf{L}} =\mathcal{F}^{\mathbf{M_1}^{-1}\mathbf{M}_1}_{\mathbf{L}}=\mathbf{I}.
\end{equation}
Given that
\begin{equation}
    (\mathcal{F}^{\mathbf{M}_1}_{\mathbf{L}})^{-1}\mathcal{F}^{\mathbf{M}_1}_{\mathbf{L}}=\mathbf{I},
\end{equation}
we immediately identify the inverse as
\begin{equation}(\mathcal{F}^{\mathbf{M}}_{\mathbf{L}})^{-1}=\mathcal{F}^{\mathbf{M}^{-1}}_{\mathbf{L}}.
\end{equation}

\subsubsection*{Proof of Property 5 (Unitarity)}
The Laplacian matrix $\mathbf{L}$ is a real symmetric matrix and thus admits an orthogonal diagonalization:
\begin{equation}
\mathbf{L}=\mathbf{U_L}\mathbf{\Lambda}\mathbf{U}^{-1}_\mathbf{L}=\mathbf{U_L}\mathbf{\Lambda}\mathbf{U}^{\mathrm{H}}_\mathbf{L},
\end{equation}
where $\mathbf{U_L}$ is a unitary matrix. 
Since the eigenvalues of a unitary matrix have unit modulus, it follows that
\begin{equation}
    \mathcal{F}^{\mathbf{M}}_{\mathbf{L}}(\mathcal{F}^{\mathbf{M}}_{\mathbf{L}})^{\mathrm{H}}=\mathbf{I}.
\end{equation}
This confirms that $\mathcal{F}^{\mathbf{M}}_{\mathbf{L}}$ is unitary.

\section{Proofs of Theorem 1}
Consider the loss function:
\begin{equation}
    \mathrm{MSE}(\mathbf{M},\mathbf{H}) = \mathbb{E}\left\{ \left\|\mathcal{F}^{\mathbf{M}^{-1}}\mathbf{H}\mathcal{F}^{\mathbf{M}}\mathbf{\tilde{f}} - \mathbf{f} \right\|_2^{2} \right\}.
\end{equation}
Since the squared norm composed with linear mappings is differentiable with respect to the operators, the MSE is differentiable with respect to $\mathcal{F},\mathbf{H}$. It suffices to prove that $\mathcal{F}$ is differentiable with respect to $\mathbf{M}$. Since the GLCT under different definitions is primarily composed of three modules GCM, GST, and GFRFT, we adopt a block-by-block argument.

\indent\textbf{Step 1: Differentiability Proof for the GCM.}
\subsubsection{GCM based on the Weighted Adjacency Matrix}
For the wAdj-GCM, the transform is defined as
\begin{equation}
    \mathcal{F}^{\xi}_\mathbf{W} = \mathbf{P_W}\mathbf{J}^{\xi}_\mathbf{W}\mathbf{P}^{-1}_\mathbf{W},
\end{equation}
where $\mathcal{CM}^\xi_{\mathbf{W}}=\mathbf{J}^{\xi}_\mathbf{W}$. As diagonal matrices commute, the chain rule yields
\[
\frac{\partial}{\partial \xi}\mathbf{J}^{\xi}_\mathbf{W}
=(\log\mathbf{J}_\mathbf{W})\,\mathbf{J}^{\xi}_\mathbf{W},
\]
Given that $\mathbf{J}^{\xi}_\mathbf{W}$ is diagonal, the derivative is evaluated entrywise. Therefore, $\mathcal{CM}_  \mathbf{W}^{\xi}$ is differentiable with respect to $\xi$.
\subsubsection{GCM based on the Laplacian Matrix}The Lap-GCM operator $\mathcal{CM}^\xi_\mathbf{L} = \operatorname{diag}(\mathbf{U_L}\hat{s}_\xi(k)),$ is differentiable in $\xi$. Its derivative computes to
\begin{equation}
    \frac{\partial}{\partial \xi}\,\mathcal{CM}^\xi_\mathbf{L}
= \operatorname{diag}\!\Big(\mathbf{U_L}\,\frac{\partial}{\partial \xi}\hat s_{\xi}\Big)
= \operatorname{diag}\!\Big(\mathbf{U_L}\big(\mathbf{i}\frac{\mathbf{\Lambda}^{2}}{\xi^2} e^{-\mathbf{i}\frac{\mathbf{\Lambda}^{2}}{\xi}})\mathbf{1}\Big),
\end{equation}
which establishes its differentiability.

\indent\textbf{Step 2: Differentiability Proof for the GST.}
\setcounter{subsubsection}{0}
\subsubsection{GST based on the Weighted Adjacency Matrix}
Let $\mathbf{S_W}(\beta)\in\mathbb{R}^{N\times N}$ be symmetric with simple spectrum, admitting the eigendecomposition
\begin{equation}
    \mathbf{S_W}(\beta)=\mathbf{V}_\sigma(\beta)\,\mathbf{\Lambda}(\beta)\,\mathbf{V}_\sigma(\beta)^\mathrm{T},
\end{equation}
where $\mathbf{V}_\sigma(\beta)=[\mathbf{v}_1(\beta),\ldots,\mathbf{v}_N(\beta)]$ is orthogonal and
$\mathbf{\Lambda}(\beta)=\mathrm{diag}(\lambda_1(\beta),\ldots,\lambda_N(\beta))$.

For each $i$,
\begin{equation}
    \mathbf{S_W}(\beta)\,\mathbf{v}_i(\beta) = \lambda_i(\beta) \mathbf{v}_i(\beta).
\end{equation}
Differentiating both sides with respect to $\beta$ 
and left-multiply by \( \mathbf{v}_j^\mathrm{T} \) with \( j \neq i \) and use orthogonality \( \mathbf{v}_j^\mathrm{T}\mathbf{v}_i = 0 \) together with \( \mathbf{v}_j^\mathrm{T} S = \lambda_j \mathbf{v}_j^\mathrm{T} \), we obtain:
\begin{equation}
    \mathbf{v}_j^\mathrm{T}(\beta) \frac{\partial\mathbf{S_W}(\beta)}{\partial\beta} \mathbf{v}_i(\beta) = (\lambda_i(\beta) - \lambda_j(\beta)) \mathbf{v}_j^\mathrm{T}(\beta) \frac{\partial \mathbf{v}_i(\beta)}{\partial\beta},
\end{equation}
which rearranges to
\begin{equation}
    \frac{\mathbf{v}_j^\mathrm{T}(\beta) \frac{\partial\mathbf{S_W}(\beta)}{\partial\beta}\,\mathbf{v}_i(\beta)}{\lambda_i(\beta) - \lambda_j(\beta)} = \mathbf{v}_j^\mathrm{T}(\beta) \frac{\partial \mathbf{v}_i(\beta)}{\partial\beta}.
\end{equation}

Since $\mathbf{v}_i$ keeps unit norm and remains orthogonal to $\{\mathbf{v}_j\}_{j\neq i}$, its derivative lies in the span of $\{\mathbf{v}_j\}_{j\neq i}$. 
\begin{equation}
\begin{split}
\frac{\partial \mathbf{v}_i(\beta)}{\partial\beta}
&= \sum_{j\neq i}\!\left(\mathbf{v}_j(\beta)^\mathrm{T}\frac{\partial \mathbf{v}_i(\beta)}{\partial\beta}\right) \mathbf{v}_j(\beta) \\
&= \sum_{j\neq i}\frac{\mathbf{v}_j^\mathrm{T}(\beta) \frac{\partial\mathbf{S_W}(\beta)}{\partial\beta}\,\mathbf{v}_i(\beta)}{\lambda_i(\beta) - \lambda_j(\beta)}\mathbf{v}_j,
\end{split}
\end{equation}
where $\mathbf{S_W}(\beta) = \frac{1}{\sigma}\mathbf{W}, \frac{\partial \mathbf{S_W}(\beta)}{\partial\beta}=-\,\frac{\mathbf{W}}{\beta^{2}}.$
This shows that under the simple-spectrum assumption, the GST basis $\mathbf{V}(\beta)$ is
differentiable in $\beta$, and its derivative can be computed from one eigendecomposition
of $\mathbf{S}_\mathbf{W}(\beta)$ together with matrix products.
\subsubsection{GST based on the Laplacian Matrix}
$\mathcal{ST}^\sigma_{\mathbf{L}} = \mathbf{S_L}$ is a diagonal matrix, differentiability with respect to 
$\sigma$ is equivalent to the differentiability of its diagonal entries. The entrywise derivative is
\begin{equation}
    \frac{\partial}{\partial \sigma} \mathbf{S}_\mathbf{L}(\sigma)
= \operatorname{diag}\!\big(-\,\varepsilon r_k\,\sigma^{-\varepsilon r_k-1}\big).
\end{equation}
\indent\textbf{Step 3: Differentiability Proof for the GFRFT.}

The GFRFT of order $\alpha$ is defined as $\mathcal{F}^{\alpha}_\mathbf{X} = \mathbf{P_X}\mathbf{J}^{\alpha}_\mathbf{X}\mathbf{P}^{-1}_\mathbf{X}.$ Since $\mathbf{P_X}$ is independent of $\alpha$, the derivative with respect to $\alpha$ is given by
\begin{equation}
    \frac{\partial}{\partial \alpha}\,\mathcal{F}_X^{\alpha}
= \mathbf{P_X}\!\left(\frac{\partial}{\partial \alpha} \mathbf{J_X}^{\alpha}\right) \mathbf{P_X}^{-1}.
\end{equation}
Therefore, the differentiability of $\mathcal{F}_\mathbf{X}^{\alpha}$ reduces to that of the diagonal matrix $\mathbf{J}_\mathbf{X}^{\alpha}$ with respect to $\alpha$.

\bibliographystyle{IEEEtran}
\bibliography{reference}

\end{document}